%% file: main.tex
\documentclass[journal]{IEEEtran}

\input{preamble}
\input{math_commands.tex}

\usepackage{tikz-cd} 

\definecolor{ReviewColor}{RGB}{166,0,116}

\title{Graph Spectral Series and Its Signatures}

\author{Feng Ji, Jingyang Dai, Apostolos Evangelidis, Xingchao Jian, Marco Rauscher, Alexander Schell%
\thanks{F.\ Ji and J.\ Dai are with the School of Electrical and Electronic Engineering, Nanyang Technological University, Singapore (Email: jifeng@ntu.edu.sg, jingyang.dai@ntu.edu.sg).}
\thanks{A.\ Evangelidis, X.\ Jian, M.\ Rauscher, and A.\ Schell are with the Technical University of Munich (Email: apostolos.evangelidis@tum.de, xingchao.jian@stsl.tu-darmstadt.de, marco.rauscher@tum.de, alexander.schell@tum.de).}
}

\begin{document}

\maketitle

\begin{abstract}
Bandlimitedness is a central modeling assumption in graph signal processing (GSP), but many graph signals exhibit substantial energy throughout the graph-frequency domain. We introduce a new framework for analyzing such non-bandlimited signals by treating their graph Fourier spectra as ordered sequences and representing them with path signatures. Specifically, motivated by the observation that non-bandlimited graph spectra often display rough, time series–like behavior, we introduce the notion of graph spectral series and develop a signature-based representation that captures global spectral structure without frequency truncation. We establish conditions under which the proposed representation is continuous and characterize its information-preserving properties. We further provide an asymptotic interpretation of graph spectral series for sequences of expanding graphs. Experiments on real-world trend-prediction and change-detection tasks show that signature features can improve performance over representations based directly on the original graph signals. The proposed approach provides a novel principled signature-based tool for analyzing non-bandlimited graph signals.
\end{abstract}

\begin{IEEEkeywords}
    Graph signal processing, graph spectral series, non-bandlimited graph signals, signature
\end{IEEEkeywords}

\section{Introduction} \label{sec:int}
Many contemporary data analysis problems involve structured data, with graphs being a prominent example. Graph signal processing (GSP) provides principled methods for analyzing signals defined on graphs, including sampling, reconstruction, and filtering \cite{Shu13, San13, Tsi15, OrtFroKov:J18, Ji19, Ji22, LeuMarMou:J23, JiaFenTay:J23}. In parallel, graph neural networks (GNNs) introduce nonlinear, trainable models specifically designed for graph-structured data and have become a central tool in modern machine learning \cite{Def16, Kip17, JiZhao25}. Both lines of work emphasize that the topology of the underlying graph can provide essential information beyond that contained in node attributes alone \cite{Seg17, Cec20, Lua21, JiJ24}.

At the core of graph signal processing (GSP) lies the notion of the \emph{graph Fourier transform (GFT)} \cite{Shu13}. It is an orthogonal transformation that maps a graph signal, represented as a vector $\bx$, into the \emph{graph frequency domain} as a vector $\widehat{\bx}$. By construction, the GFT encodes the underlying graph structure, as its transformation matrix is formed from an eigenbasis of a prescribed graph operator. The resulting graph spectrum $\widehat{\bx}$ characterizes the signal’s response across different levels of smoothness on the graph. Therefore, most conveniently, $\widehat{\bx}$ can be arranged as an ordered sequence.

In the GSP literature, ``bandlimitedness'' is a commonly used notion for characterizing graph signals \cite{Shu13, Tsi15, Puy18, OrtFroKov:J18, Ort22}. Intuitively, a graph signal $\bx$ is \emph{bandlimited} if most of its energy is concentrated in only a small number of GFT components. In many studies, these components are assumed to lie in the low-frequency range, reflecting the idea that informative graph signals tend to vary smoothly over the graph. As a result, high-frequency components, which correspond to rapid changes across neighboring nodes, are typically small or negligible.

\begin{figure}
    \centering
    \includegraphics[width=1\linewidth]{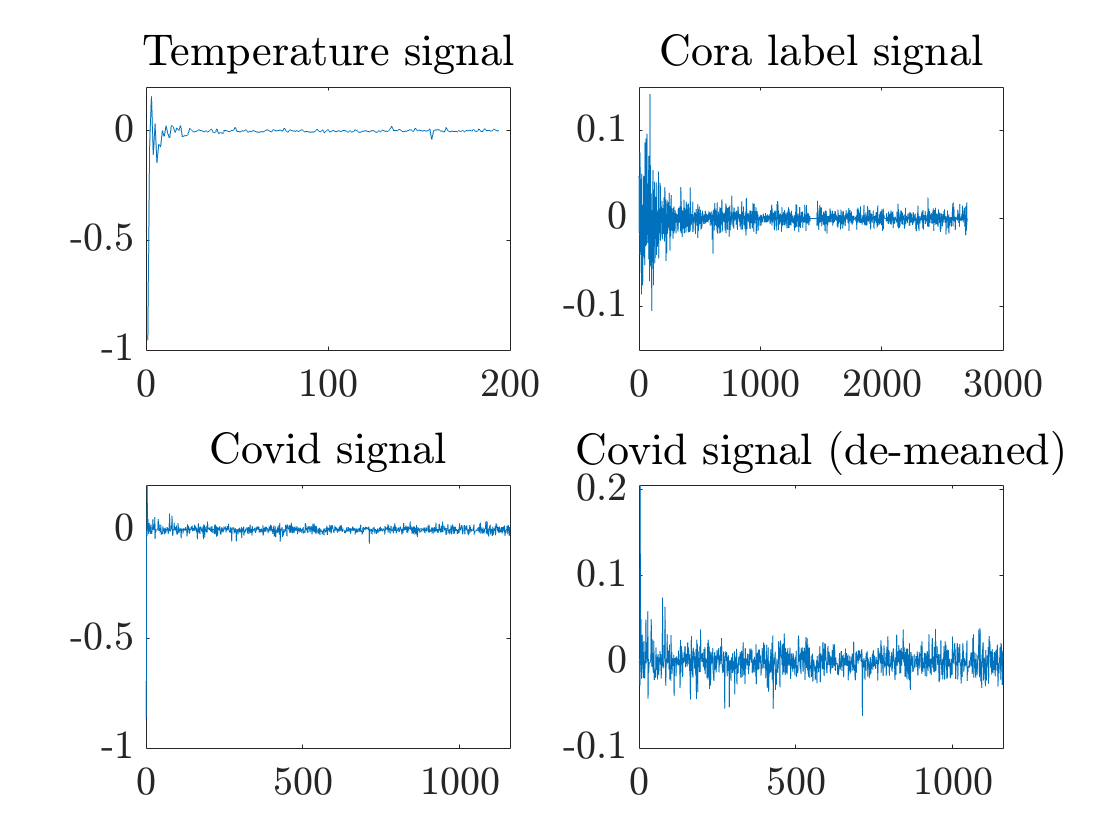}
    \caption{We present several examples of graph signal spectra, where the amplitude is plotted against the frequency index. The signals considered include daily temperature readings from U.S. weather stations, the label signal on the Cora citation graph, and the number of 20-day Covid-19 infection cases in California counties (with a spatio-temporal graph). We observe that smooth signals, such as temperature readings, may exhibit approximately bandlimited behavior, whereas other signals can spread their energy across the entire frequency spectrum.}
    \label{fig:wse}
\end{figure}

However, in practice, the bandlimited assumption can be far from realistic. As illustrated in \cref{fig:wse}, the frequency responses of graph signals may exhibit non-negligible energy across the entire frequency domain. In such cases, overly simplifying a signal as bandlimited can lead to misrepresentation and misinterpretation. In some scenarios, a graph signal may appear to have a single dominant leading component under a Laplacian-based GFT. This phenomenon often reflects a large global average relative to local variations, rather than genuine smoothness across the graph (e.g., compare the original and de-meaned Covid-19 signals in \cref{fig:wse}). Analyzing such signals under a bandlimited assumption may therefore yield oversimplified or even misleading conclusions. If the signal is first preprocessed by removing its mean, the resulting spectral components associated with local variation become more visible. They can reveal additional fine-grained and informative structures that would otherwise be obscured. 

In summary, while the bandlimited assumption has proven useful in many settings, it is equally important to be able to analyze graph signals that do not satisfy this assumption, when frequency coefficients display substantial local variation and complex global structure. In this regard, our key observation is that the spectrum of a non-bandlimited graph signal often resembles a time series indexed by graph frequency, resembling those arising in financial data. In particular, the spectrum may exhibit an ``oscillatory'' or ``rough'' profile with pronounced local fluctuations (cf.\ \cref{fig:wse}). Such behavior motivates using signatures \cite{Fri12, Lyo14} from rough path theory. The finite-graph interpolants used here are continuous bounded-variation paths, so their signatures are defined by classical Riemann-Stieltjes integration. 

The above observation motivates us to adopt signature theory \cite{Fri12, Lyo14, Lev16, Che26} as a tool for analyzing graph signals without explicitly imposing a bandlimited assumption. Signature theory provides a framework for analyzing continuous-time series, including those with rough trajectories. Intuitively, it characterizes a path through collections of iterated integrals, which serve as expressive features of the time series. Since integration naturally aggregates information along a path, signature features are well-suited for capturing the global structure of time series associated with a graph signal spectrum. This perspective is particularly valuable when inspecting a small number of sampled frequency responses is insufficient, as strong local fluctuations may obscure meaningful representations.

The rest of the paper is organized as follows. In \cref{sec:pre}, we review preliminary material on graph signal processing and signature theory. In \cref{sec:tsp}, we develop the proposed framework for using signatures to analyze graph signals. Since signature theory applies naturally to time series, we provide a theoretical justification for interpreting graph signal spectra as time series in \cref{sec:cog}. Numerical experiments demonstrating the effectiveness of the proposed framework are presented in \cref{sec:nr}, and concluding remarks are given in \cref{sec:con}. All proofs are deferred to the appendices.

Our \emph{main contributions} are summarized as follows:
\begin{itemize}
\item We introduce the notion of graph spectral series and develop a novel framework, grounded in signature theory, for analyzing graph signals that are not necessarily bandlimited. For signed spectra in a fixed GFT basis, we establish recoverability up to the stated ambiguity and linear approximation on suitable compact signal sets.
\item We provide convergence results for graph spectral series along sequences of expanding graphs, thereby giving a theoretical basis for their interpretation as time series and providing an asymptotic path interpretation on every proper subinterval of the frequency-index domain.
\item We evaluate the proposed features on real-world trend-prediction and change point detection tasks, and show that they can improve performance relative to using the original graph signals directly.
\end{itemize}

\paragraph*{Scope of the work} This work introduces a novel signature-based tool for analyzing graph signals, particularly those that are not well described by bandlimited models. It is written primarily for the \emph{GSP community}, with the goal of making signature methods accessible for graph-signal analysis. Accordingly, we present the necessary signature-theoretic background and establish conditions under which graph spectra can be interpreted as ordered paths. Although the underlying tools originate in rough-path theory, their specialization to graph spectral series clarifies their role and applicability in GSP. This work complements other applications of signature
methods in signal processing, including their use in nonlinear independent component analysis \cite{SchOb23} and in quantifying the robustness of blind source separation \cite{sch23}. More broadly, the connection established in this work opens new avenues for applying signature theory to seemingly less related fields in signal processing.

\section{Preliminaries} \label{sec:pre}
This section presents a concise overview of graph signal processing (GSP) and signature theory, with emphasis on the essential concepts required for the developments that follow.

\subsection{Graph Fourier transform and signal spectrum} \label{sec:gft}

Suppose $G=(V,E)$ is an undirected graph of size $|V|=n$. Let $\bS_n$ be a \emph{graph shift operator}, which is an $n\times n$ symmetric matrix. Common choices of $\bS_n$ include the adjacency matrix $\bA_n$, the Laplacian $\bL_n$ or their normalized versions $\overline{\bA_n}, \overline{\bL_n}$. Being symmetric, $\bS_n$ admits an \emph{orthonormal eigenbasis} $\{\bu_1, \ldots, \bu_n\}$. They are ordered (increasingly) according to their associated eigenvalues $\lambda_1 \leq \lambda_2 \leq \ldots \leq \lambda_n$. 

A \emph{graph signal} $\bx \in \mathbb{R}^n$ assigns a number $\bx(i)$ to each node $v_i\in V$. At the core of Fourier theory lies the idea of analyzing signals by representing them in the frequency domain. For graph signal, this is achieved via the \emph{graph Fourier transform (GFT)}, defined for $\bx$ as $\widehat{\bx} \in \mathbb{R}^n$: $\widehat{\bx}(i) = \langle \bu_i, \bx \rangle$.

As noted in \cref{sec:int}, numerous works impose the \emph{bandlimited} assumption on $\bx$, i.e., $\widehat{\bx}(i) = 0$ for $i\geq n_0$, where $n_0 \ll n$. Such a characterization is often an oversimplification for real-world signals (as illustrated in \cref{fig:wse}). This paper aims to develop a generic framework for systematically characterizing graph signals and their Fourier transforms.

Our main \emph{insight} is to leverage the intrinsic ordering of graph frequencies, thereby interpreting the transformed signal $\widehat{\bx}$ as a ``time-ordered sequence''. This viewpoint enables the systematic application of tools from signature theory, as detailed in the next subsection.

\subsection{Signature of time series} \label{sec:sot}

The signature of a continuous time series is a sequence of coefficients that encodes many of its fundamental analytic and geometric properties. Defined iteratively through path integrals, the signature provides a principled representation of the global structure of a time series. This representation is particularly effective for ``rough'' time series exhibiting pronounced and irregular fluctuations. Our overview follows largely from \cite{Che26}.

Before formally introducing the concept of signature, consider a closed interval $[a,b]$. For a \emph{continuous} function $f$ on $[a,b]$, its \emph{total variation} (cf.\ \cite{Rud76}) is defined as
\begin{align*}
    V(f;[a,b]) = \sup_{\by = (y_1,\ldots,y_l)}\sum_{i=1}^{l-1}|f(y_{i+1})-f(y_i)|, 
\end{align*}
where the supremum runs over finite sequences $\by$ of points $y_1=a< y_2 <\ldots< y_{l-1}< y_l=b$. The function $f$ is of \emph{bounded variation (BV)} if $V(f;[a,b])< \infty$. 

We denote the set of \emph{continuous BV functions} on $[a,b]$ by $BV([a,b])$. They are used to define path integrals via the \emph{Riemann–Stieltjes integral} \cite{Rud76}. Specifically, for $f \in BV([a,b])$ and $g$ continuous, the following path integral is well-defined for any $t\in [a,b]$
\begin{align} \label{eq:igf}
    \int_a^t \!g\ud f = \lim_{\Delta\by \to 0} \sum_{1\leq i\leq l-1} g(y_i)\big( f(y_{i+1})-f(y_i) \big),
\end{align}
where $\Delta\by$, for the increasing sequence $\by=(y_1=a,y_2,\ldots,y_l=t)$, is $\max\{y_{i+1}-y_i, 1\leq i\leq l-1\}$.

We are ready to introduce the notion of signature. The idea is to take path integrals recursively in the form of \cref{eq:igf}. For the \emph{base case}, for any $f \in BV([a,b])$, we define:
\begin{align*}
   S_f^{\emptyset}(t)=1; \text{ and } S_f^1(t) = \int_a^t\!\ud f.
\end{align*}

In general, consider any finite ordered set of functions $\boldf = (f_1,\ldots,f_d)$, with $f_i\in BV([a,b]), 1\leq i\leq d$. Its signature is a countable sequence of numbers defined as follows.

\begin{Definition} \label{def:ljj}
Let $J = (j_1, \ldots, j_k)$ with each $j_i\leq d$ be a multi-index set of \emph{length} $|J|=k$. Denote its sub-index set $(j_1,\ldots, j_{k-1})$ of length $k-1$ by $J_-$. We define the function $S_{\boldf}^J$ (on $[a,b]$) with the following recursive formula:
\begin{align*}
    S_{\boldf}^J(t) = \int_{a}^t\!S_{\boldf}^{J_-}(s)\ud f_{j_k}(s), \text{ and } S_{\boldf}^{\emptyset}(t)=1. 
\end{align*}
The \emph{signatures} $\bs_{\boldf} = (s_{\boldf}^J)_J \in \mathbb{R}^{\infty}$ of the functions $\boldf$ is the countable collection of numbers
\begin{align*}
    s_{\boldf}^J = S_{\boldf}^J(b),
\end{align*}
over all multi-indices $J$ (including $J=\emptyset$).
\end{Definition}

Although the signature $\bs_{\boldf}$ of a collection of functions $\boldf$ is defined recursively through path integrals, efficient algorithms and software implementations exist for its computation,\footnote{\url{https://pypi.org/project/esig/}.} which we employ whenever possible. We analyze a few special cases in the next example.

\begin{Example} \label{eg:fal}
For a length $1$ index $j\leq d$, we have $s_{\boldf}^{j} = f_j(b)-f_j(a)$. This follows directly from \cref{eq:igf}.  

Consider $J$ of length $2$. For $J=(j,j)$, we have $s_{\boldf}^J = (f_j(b)-f_j(a))^2/2$ by integration by parts. If $J = (j_1,j_2)$ and $J' = (j_2,j_1)$ with $j_1\neq j_2\leq d$, then $s_{\boldf}^J, s_{\boldf}^{J'}$ satisfy the identity $s_{\boldf}^J-s_{\boldf}^{J'} = 2A$, where $A$ is the Levy area (see \cref{fig:tla} for an illustration) enclosed by the curve $(f_{j_1}(t),f_{j_2}(t)), t\in [a,b]$ and its chord in $\mathbb{R}^2$ \cite{Lev40, Che26}.    

The \emph{takeaway} is that for $J$ with small length, $s_{\boldf}^J$ often encodes global geometric information of the paths in $\boldf$. 
\begin{figure}[!htb]
    \centering
    \includegraphics[width=0.7\linewidth]{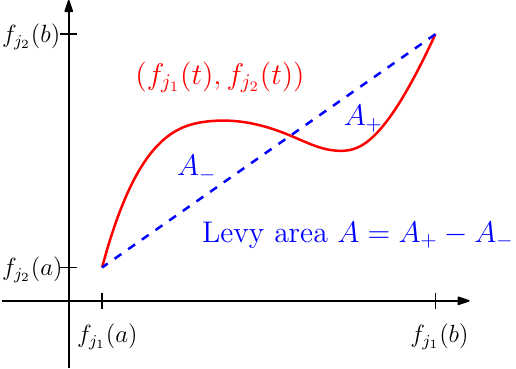}
    \caption{The figure illustrates the Levy area $A$ of the curve $(f_{j_1}(t),f_{j_2}(t)), t\in [a,b]$. The diagonal chord and the curve bound a positive and a negative region, whose areas are $A_+$ and $A_-$, respectively. The Levy area is $A = A_+-A_-$.} \label{fig:tla}
\end{figure}
\end{Example}

In practice, the infinite signature sequence $\bs_{\boldf}$ can be truncated to yield a ``finite feature representation'': $\bs_{\boldf,\eta} = (s_{\boldf}^J)_{|J|\leq \eta}$, for a chosen \emph{length threshold} $\eta$. A truncated sequence $\bs_{\boldf,\eta}$ always includes $s_{\boldf}^J$ for small $|J|$, which prioritizes capturing global features of $\boldf$ (cf.\ \cref{eg:fal}). This is particularly useful when $\boldf$ includes continuous time series, whose path is oscillatory with many local fluctuations. Such a phenomenon is common for the graph spectral series introduced in the next section. 

For readers' convenience, we include a list of notations for key concepts in \cref{tab:nl}, including those defined in the next section.  
\begin{table}[!htb]
\renewcommand{\arraystretch}{1.5} 
    \caption{Notation list in chronological order.}
    \label{tab:nl}
    \centering
    \begin{tabular}{cc}
    \hline
       $V(f;[a,b])$  & Total variation of $f$ on $[a,b]$ \\ 
           \hline
           $\boldf$ & A finite ordered set of functions on $[a,b]$
           \\
           \hline
        $S_{\boldf}^J(\cdot)$ & Signature function of $\boldf$ given multi-index $J$ \\
         \hline
        $s_{\boldf}^J$ & Signature of $\boldf$ given multi-index $J$
        \\
        \hline
        $s_{\boldf}$ & Signatures of $\boldf$ over all multi-indices \\
         \hline
         $\bs_{\boldf,\eta}$ & Signatures of $\boldf$ with multi-index length $\leq \eta$   \\
         \hline
        $p(I;\bz)(\cdot)$ & Interpolation of $\bz = (z_1,z_2)$ on $I$ via $p$ \\
        \hline
        $p(\mathcal{P}_n; \widehat{\bx})(\cdot)$ & Graph spectral series associated with signal $\bx$\\
        \hline
        $\bX$ & An ordered set of graph signals \\
        \hline
        $f_j, j\leq d$ & Graph spectral series associated with the signal $\bx_j$\\
        \hline
        $f_{d+1}(\cdot)$ & The identity function $f_{d+1}(t)=t$\\
        \hline
         $\boldf_{\bX}$ & The set of functions $(f_j)_{j\leq d+1}$ derived from $\bX$ \\
        \hline
         $\bs_{\bX}$ & Signatures of $\bX$  (i.e., $\bs_{\boldf_{\bX}}$ associated with $\boldf_\bX$)\\
         \hline
         $\bs_{\bX,\eta}$& $\eta$-truncated signatures of $\bX$ (i.e., $\bs_{\boldf_{\bX},\eta}$) \\
         \hline
         $\bs_{\bx,\eta}$ & $\eta$-truncated signatures if $\bX = \{\bx\}$ \\
         \hline
    \end{tabular}
\end{table}

\section{The signature perspective of graph signal processing} \label{sec:tsp}

In this section, we introduce the notion of graph spectral series and explain how to use the signature to analyze graph signals via graph spectral series. 

\subsection{Graph spectral series and its signature} \label{sec:gss}

Following \cref{sec:gft}, we have a graph $G=(V,E)$ of size $n\geq 2$, and a pre-determined graph shift operator $\bS_n$. For a graph signal $\bx$ on $G$, its GFT is $\widehat{\bx}$. To apply the signature tool, we first convert $\widehat{\bx}$ into a continuous time series on $[0,1]$.\footnote{In principle, any interval $[a,b]$ can be used, as simple translation and scaling convert it into $[0,1]$.}  

We assume there is a \emph{partition} $\mathcal{P}_n = \{I_{1,n}, \ldots, I_{n-1,n}\}$ of $[0,1]$ into $n-1$ sub-intervals. Each $I_{i,n}, i\leq n-1$ is of the form $[e_{i,n},e_{i+1,n}]$, and the end-points satisfy $e_{1,n}=0 < e_{2,n} < \ldots < e_{n,n}=1$.  

Let $p$ be an \emph{interpolation function}, i.e., given any interval $I=[a,b]$ and $\bz= (z_1, z_2)$, then $p(I;\bz)$ is a function on $I$ such that $p(I;\bz)(a) = z_1$ and  $p(I;\bz)(b) = z_2$. A familiar example is the \emph{linear interpolation} such that $p(I;\bz)(t) = z_1+ (z_2-z_1)(t-a)/(b-a)$.

\begin{Definition} \label{defn:gpa}
    Given $\mathcal{P}_n$ and $p$, the \emph{signed graph spectral series} associated with the graph signal $\bx$ is the function $p(\mathcal{P}_n; \widehat{\bx})$ on $[0,1]$, which is an extension of the interpolation $p$ (above) defined as follows: if $t \in I_{i,n}\subset [0,1]$, then 
    \begin{align*}
    p(\mathcal{P}_n; \widehat{\bx})(t) = p\Big(I_{i,n}; \big(\widehat{\bx}(i),\widehat{\bx}(i+1)\big)\Big)(t).
    \end{align*}
Let $|\widehat{\bx}|$ be the sequence of numbers $\big(|\widehat{\bx}(1)|,\ldots,|\widehat{\bx}(n)|\big)$. The \emph{absolute spectral series} associated with $\bx$ is $p(\mathcal{P}_n; |\widehat{\bx}|)$. 
\end{Definition}

\begin{Remark}
Note that if $\bu_i$ is an eigenvector of $\bS_n$, then so is $-\bu_i$. Consequently, $\widehat{\bx}$ depends on the arbitrary choice of the sign (direction) of each eigenvector. When analyzing multiple signals defined on the same graph, this ambiguity is inconsequential, and one may fix the direction of each eigenvector without loss of generality. However, when comparing signals across different graphs (see \cref{sec:cog}), the corresponding shift operators generally differ, making it difficult to consistently align eigenvector directions. In this setting, the absolute spectral series becomes particularly useful, as it is invariant to eigenvector sign ambiguities. We omit ``signed'' or ``absolute'' only when the choice is clear.
\end{Remark}

The natural next step is to incorporate the signature theory discussed in \cref{sec:sot}. We consider the slightly more general setting when multiple graph signals are considered simultaneously.

Fix the partition $\mathcal{P}_n$ and the interpolation method $p(\cdot;\cdot)$. Let $\bX = (\bx_1,\ldots,\bx_d) \in \mathbb{R}^{n\times d}$ be an ordered set of graph signals. Assume we consider either the graph spectral series or the absolute graph spectral series, consistently for all signals in $\bX$. For convenience, we accordingly use $f_j$ to denote either $p(\mathcal{P}_n; \widehat{\bx_j})$ or $p(\mathcal{P}_n; |\widehat{\bx_j}|)$, for $j\leq d$. 

Suppose each $f_j \in BV([0,1])$ for $j\leq d$ (see \cref{sec:taf} below for discussions on this condition). Let $f_{d+1}(t)=t$ be the identity function on $[0,1]$, which clearly belongs to $BV([0,1])$. By collecting all functions, we form\footnote{There are benefits for introducing $f_{d+1}$. For example, if $d=1$, then the signature of $f_1$ is just the sequence consisting of $(f_1(b)-f_1(a))^j/j!$. Therefore, we need to augment the identity function (as the ``time'' component) to make the resulting signature more informative. }  
\begin{align} \label{eq:bbf}
    \boldf_{\bX} = (f_1,\ldots,f_d, f_{d+1}) \in BV([0,1])^{d+1}. 
\end{align}

\begin{Definition} \label{defn:tsb}
The \emph{signature (feature)} $\bs_{\bX}$ of $\bX$ is defined as 
\begin{align*}
    \bs_{\bX} = \bs_{\boldf_{\bX}} =(s_{\boldf_{\bX}}^J)_J \in \mathbb{R}^{\infty},
\end{align*}
where $\bs_{\boldf_{\bX}}$ is introduced in \cref{def:ljj}. Similarly, for any $\eta>0$, the truncated signature (feature) $\bs_{\bX,\eta}$ is defined as
\begin{align*}
    \bs_{\bX,\eta} = \bs_{\boldf_{\bX},\eta} =(s_{\boldf_{\bX}}^J)_{|J|\leq \eta} \in \mathbb{R}^{D_{\eta}},
\end{align*}
where $D_{\eta}$ is the number of $J$ whose length is bounded by $\eta$.
\end{Definition}

\paragraph*{Convention} For convenience, if $\bX = \{\bx\}$ for a single graph signal, we also use $\bs_{\bx}$ (resp.\ $\bs_{\bx,\eta}$) for $\bs_{\bX}$ (resp.\ $\bs_{\bX,\eta}$).

Next, we discuss fundamental properties of the graph signal signature. From an application perspective, we regard the signature or its truncated version as a \emph{``feature representation''} of graph signals, which can be readily employed in standard downstream tasks, as outlined also in the next subsection.

\subsection{Characterizing signals and graphs}

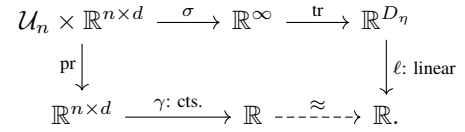
\begin{figure}[!htb]
\centering
\begin{tikzcd}
\mathcal{U}_n\times \mathbb{R}^{n\times d} \arrow{r}{\sigma} \arrow[swap]{d}{\text{pr}} & \mathbb{R}^{\infty} \arrow{r}{\text{tr}} & \mathbb{R}^{D_{\eta}} \arrow{d}{\ell\text{: linear}} \\%
\mathbb{R}^{n\times d} \arrow{r}{\gamma\text{: cts.}}& \mathbb{R} \arrow[dashed]{r}{\approx} & \mathbb{R}.
\end{tikzcd}
\caption{The signature framework, with $\mathcal{U}_n\subseteq O(n)$ the space of chosen ordered GFT bases. The maps ``pr'' and ``tr'' are projection to the signals and signature truncation. Theorem~\ref{thm:sbi} gives finite ambiguity for unit-norm signals at fixed $\bU$, or for $\bU$ at $n$ fixed independent probes. After fixing $\bU$ and restricting signals to a compact $K$ satisfying \cref{thm:sti}(b), the two routes approximate the same continuous function.
} \label{fig:tsd}
\end{figure}

The primary goal of this subsection is to explain how the signature of graph spectral series can be used in practice. We begin by discussing its theoretical properties, which provide the necessary motivation and intuition for its practical application. A schematic diagram for illustration is given in \cref{fig:tsd}.

We now state our main results for signatures derived from signed graph spectral series.

\begin{Theorem}[Signal and GFT recoverability] \label{thm:sbi}
\begin{enumerate}[(a)]
    \item Suppose $\bx$ is a unit-norm signal on $G$. Then there exists at most one other unit-norm signal whose signature equals $\bs_{\bx}$.
    \item Let $\bX = \{\bx_1,\ldots,\bx_n\}$ for $n$ linearly independent (signal) vectors. Then there are at most finitely many GFTs that yield the same signature $\bs_{\bX}$. 
\end{enumerate}    
\end{Theorem}

\begin{Theorem}[Linear approximation] \label{thm:sti}
Suppose the interpolation $p(\cdot; \cdot)$ is strongly consistent.\footnote{Consistency conditions are defined in \cref{sec:taf} b), where we present a systematic discussion on different conditions of the interpolation function.} Then the following holds:
\begin{enumerate}[(a)]
    \item Let $\mathbb{R}^{n\times d}$ be the space of ordered sets of $d$ graph signals, i.e., consisting of $\bX = (\bx_1,\ldots,\bx_d)$. For any $\eta\geq 0$, the map $\sigma_{\eta}: \mathbb{R}^{n\times d} \to \mathbb{R}^{D_{\eta}}, \bX \mapsto \bs_{\bX,\eta}$ is continuous.
    \item Let $K$ be a compact subset of $\mathbb{R}^{n\times d}$ such that the following holds: Given distinct $\bX = (\bx_1,\ldots,\bx_d), \bX' = (\bx_1',\ldots,\bx_d') \in K$, then there is at least a $j\leq d$ such that for any $c_j\in \mathbb{R}$
    \begin{align*}
        \widehat{\bx_j}-\widehat{\bx_j'} \neq c_j\mathbf{1}, \quad \text{for } \ \mathbf{1}=(1,\ldots,1)^\intercal\in\R^n.
    \end{align*}
    For any continuous function $\gamma: K \to \mathbb{R}$ and $\epsilon>0$, there is a $\eta\geq 0$ and a \emph{linear} function $\ell: \mathbb{R}^{D_{\eta}} \to \mathbb{R}$ such that: for any $\bX\in K$, we have the following approximation for the truncated signature $\sigma_{\eta}(\bX)$:   
    \begin{align*}
        |\gamma(\bX) - \ell\circ \sigma_{\eta}(\bX) |\leq \epsilon.
    \end{align*}
\end{enumerate}
\end{Theorem}

\paragraph*{Intuitions and implications} For a fixed GFT basis and signed spectral series, \cref{thm:sbi}(a) asserts that the signature ``almost'' uniquely determines a unit-norm graph signal.\footnote{This ambiguity arises from the technical fact that two functions differing by a constant yield identical path integrals.} Similarly, \cref{thm:sbi}(b) indicates that $n$ linearly independent vectors can be used to probe and characterize the underlying Fourier transform operator (see \cref{eg:tst} below for a graph-classification application). Taken together, these results suggest that the signature can serve as a defining feature, up to these ambiguities, for either the normalized signal itself or the graph's chosen Fourier basis.

In signal processing and machine learning, input signals are typically transformed through nonlinear feature maps $\gamma(\cdot)$, e.g., MLPs, to obtain representations suitable for downstream tasks. \cref{thm:sti}(b) places an additional separation condition on $K$. For a finite collection of independently sampled inputs with densities, this condition holds almost surely: no two distinct samples have Fourier transforms differing only by channel-wise additive constants. The condition thus rules out potential ambiguities associated with the signature's translation invariance. \cref{thm:sti} shows that the signature can be viewed as a form of ``universal feature'', in the sense that any continuous real-valued feature map $\gamma(\cdot)$ defined on $K$ can be approximated uniformly, to arbitrary accuracy, by a linear function $\ell(\cdot)$ acting on a finite subset of the signature. The same conclusion applies componentwise to finite-dimensional feature maps. Since graph signals can be conveniently converted into signatures as a preprocessing step (via $\sigma_{\eta}(\cdot)$), it is therefore natural and practically appealing to adopt the signature as a signal representation.

\begin{Example} \label{eg:tst}
\cref{thm:sbi}(b) suggests that signal signatures give substantial information about the GFT and hence the graph itself; and thus they can be potentially used for graph classification. To give empirical evidence, we consider standard biochemical graph classification datasets MUTAG and PROTEIN \cite{Mor20}. For demonstration purposes, we implement a simple signature-based algorithm: we apply \cref{defn:tsb} to the vector of node features, and the resulting signature features are passed to a simple classifier, e.g., LightGBM \cite{Mou25}. 

The resulting test accuracy is $87.2\%$ on MUTAG and $73.8\%$ on PROTEIN, which is comparable to more sophisticated GNN models and graph-kernel methods such as the WL subtree kernel ($85.6\%, 72.5\%$) and WL graph kernel ($86.9\%, 74.3\%$) \cite{Ngu21}. The observation aligns with \cref{thm:sbi}(b).
\end{Example}

\section{Convergence of graph spectral series: a conceptual foundation} \label{sec:cog}

In the existing literature, signature theory is primarily developed for time series. By contrast, at first glance, the graph spectral series introduced in \cref{sec:gss} does not explicitly incorporate a temporal component. In this section, we further develop the theoretical foundations of graph spectral series and provide an intuitive interpretation that relates it to a time series, viewed through the lens of randomly expanding graphs. While this section plays a crucial role in the theoretical development of the proposed framework, readers primarily interested in practical applications may choose to skip the technical details in \cref{sec:taf}.

\subsection{The intuitive picture: random expanding graphs} \label{sec:tir}
\begin{figure}
    \centering
    \includegraphics[width=0.8\linewidth]{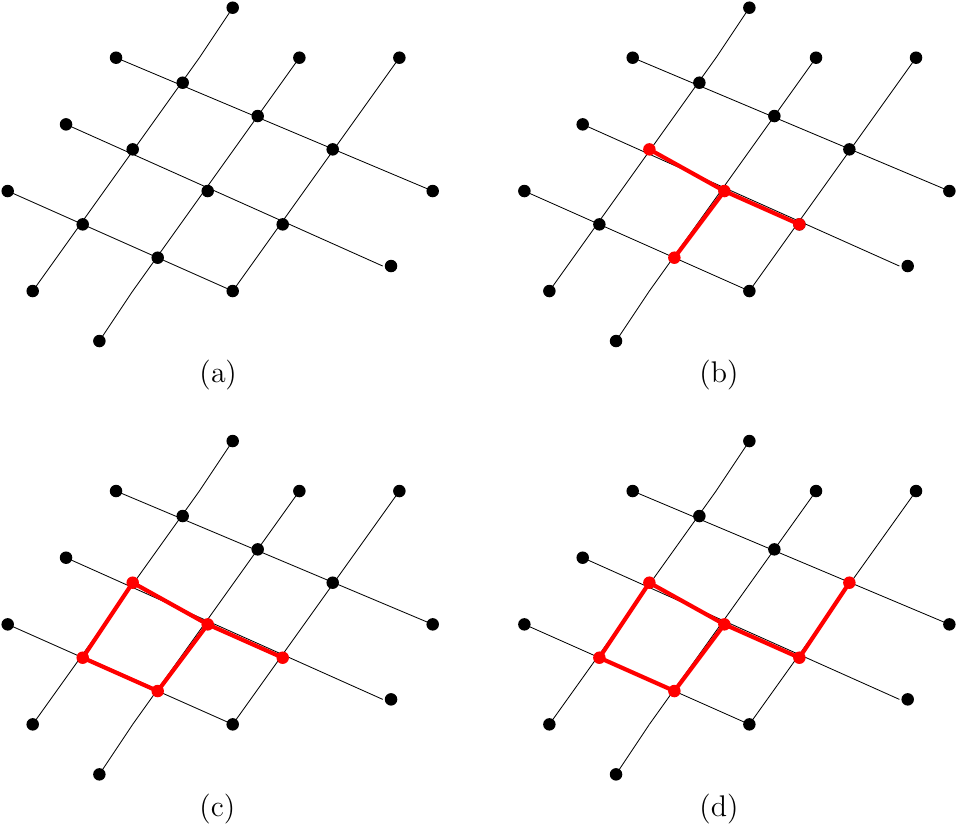}
    \caption{A ``random expansion'' on the ambient graph $\widehat{G}$ in (a). In each iteration, the (red) subgraph expands to include a random neighbor.}
    \label{fig:res}
\end{figure}
In this subsection, we provide intuition for interpreting the graph spectral series as a time series. Our discussion is motivated by the analogy that a random walk gives rise to the continuous time series of Brownian motion \cite{Dud99}.

Consider the following scenario. There is a large ambient graph $\widehat{G}$. Let $G_1$ be the trivial graph on a node $v_1$ in $\widehat{G}$. We model the time evolution of graphs starting from $G_1$ as follows: at each discrete time-stamp $n$, we expand the graph $G_{n-1}$ to form $G_n$ by including a randomly chosen neighbor (cf.\ \cite{Per14}) of $G_{n-1}$ in $\widehat{G}$ (see an illustration in \cref{fig:res}). 

Suppose there is a signal $\bx$ on $\widehat{G}$. Then by considering the restriction of $\bx$ to $G_n$, we obtain the signal $\bx_n$ on $G_n$. Notice we have the obvious \emph{consistency} that $\bx_{n-1}$ and $\bx_n$ agree on $n-1$ components. Thus, viewing $\bx_{n-1}$ as a time series allows $\bx_n$ to be interpreted as the extension of that series by one additional time component.

However, this perspective has a serious limitation: $\bx_n$ itself does not encode any graph information. This issue is alleviated by considering its graph Fourier transform (GFT) $\widehat{\bx_n}$. On the other hand, $\widehat{\bx_{n-1}}$ and $\widehat{\bx_n}$ may differ across all components, which prevents $(\widehat{\bx_n})_{n \ge 1}$ from forming a consistent family of time series. The main objective of this section is to address this issue by showing they ``converge'' in an appropriate sense as $n \to \infty$.

We rigorously formulate the above setup in \cref{sec:taf}. In particular, the graphon theory \cite{Lov:12, Jan:J13, Rui20, Lua21, Shu25} can be useful for modeling infinite graphs, which shall be discussed in \cref{sec:taf} d).

\subsection{The analytical framework} \label{sec:taf}
\paragraph{The formal setup} In general, we assume there is a sequence of graphs $(G_n)_{n\geq 1}$, each associated with a shift operator $\bS_n$. Recall that for each $n$, there is a graph signal $\bx_n$ on $G_n$, whose GFT is $\widehat{\bx_n}$. For $n\geq 2$, there is a partition $\mathcal{P}_n=\{I_{1,n},\ldots,I_{n-1,n}\}$ of $[0,1]$. With a fixed interpolation $p(\cdot;\cdot)$, we are interested in the following:
\begin{Problem}
Whether the absolute spectral series 
\begin{align} \label{eq:fmm}
f_n = p(\mathcal{P}_n; |\widehat{\bx_n}|/\gamma_n),
\end{align}
for suitable normalization factors $\gamma_n$, converges as $n\to\infty$. 
\end{Problem}

We cannot expect convergence to hold unconditionally. Instead, mild conditions must be imposed on the primary contributors of $f_n$, namely: $p(\cdot;\cdot), \mathcal{P}_n$ and $|\widehat{\bx_n}|$. These conditions are discussed separately in parts b)–d) below.

\paragraph{The interpolation function}
The interpolation is \emph{Lipschitz continuous} if for any $I=[a,b]$ and $\bz = (z_1,z_2)$, the interpolation function $f=p(I;\bz)$ (cf.\ \cref{sec:gss}) is Lipschitz continuous on $[a,b]$; and hence, $f\in BV([a,b])$. 

For the interpolation to be \emph{regular}, we require that the Lipschitz constant of $f$ can be chosen to take the form 
\begin{align*}
\frac{g(|f(a)-f(b)|)}{b-a} = \frac{g(|z_1-z_2|)}{b-a},
\end{align*}
for some monotone increasing function $g$ that does not depend on the interval $I$. The factor $1/(b-a)$ reflects the natural scaling of interpolation: traversing the same increment over a shorter interval requires a proportionally steeper interpolant.

For the interpolation to be \emph{consistent}, we require that there is a single continuous function $\phi: \mathbb{R}^2\times [0,1]\to \mathbb{R}$, with $\phi(z_1,z_2,0)=z_1$ and $\phi(z_1,z_2,1)=z_2$, such that for every $I = [a,b]$ and $\bz$,
\begin{align} \label{eq:pff}
f(y) = \phi\Big(z_1,z_2,\frac{y-a}{b-a}\Big), \quad y\in I,
\end{align}
i.e., the interpolant depends on the interval only through the \emph{relative position} of $y$ within $I$. The interpolation is \emph{strongly consistent} if $\phi$ is continuously differentiable.

\emph{Intuitively}, both regularity and consistency require that the continuity behavior of $f$ on $[a,b]$ is controlled by its values at the endpoints of $[a,b]$, uniformly over intervals. For example, the linear interpolation is both regular and strongly consistent.

\paragraph{Interval partitions}
For the partitions, we are interested in their asymptotic properties as $n\to \infty$, as this addresses the convergence question. 

The set of partitions $\mathcal{P} = \{\mathcal{P}_n, n\geq 2\}$ is said to have \emph{diminishing tails} if 
\begin{align*}
    \lim_{N\to \infty} \limsup_{n\to \infty}\sum_{i=N}^{n-1}|I_{i,n}| = 0.
\end{align*}
This condition requires that, asymptotically, a finite initial block of intervals carry arbitrarily much of the interval's total length.

On the other hand, we say $\mathcal{P}$ satisfies the \emph{non-vanishing} condition if for each $i$, $\liminf_{n\to \infty} |I_{i,n}| > 0$. If for each $i$, $\lim_{n\to \infty} |I_{i,n}|$ exists, then the set of partitions is called \emph{index-wise convergent}. The non-vanishing condition ensures that none of the intervals are asymptotically trivial.

\paragraph{Graph sequence models}
The motivational model in \cref{sec:tir} can be more rigorously described as follows. $\widehat{G} = (\widehat{V},\widehat{E})$ is a countably infinite graph such that for each $v\in \widehat{V}$, its degree is finite. Starting from $G_1 = \{v_1\}$ in an infinite component of $\widehat{G}$, for $n>1$, $G_n = (V_n,E_n)$ has the node set $V_n=V_{n-1}\cup \{v_n\}$, where $v_n$ is chosen uniformly from the outer boundary $\{v\notin V_{n-1}: v\text{ is adjacent to some }w\in V_{n-1}\}$. This set is finite and nonempty by local finiteness and the infinite-component assumption. Accordingly, $E_n$ contains $E_{n-1}$ and all edges connecting $v_n$ and $V_{n-1}$. 

A signal $\bx \in \mathbb{R}^{\infty}$ on $\widehat{G}$ assigns a value to each node $v\in \widehat{G}$. Therefore, its restriction on $V_n$ induces a graph signal $\bx_n$ on $G_n$. As we do not impose any metrical condition on $\bx$, an appropriate scaling factor $\gamma_n$ in \cref{eq:fmm} is needed to prevent $\bx_n$ from ``exploding'' as $n\to \infty$.

We call the above procedure the \emph{random expansion model}, which warrants subsequential convergence as in \cref{thm:tso}. 

To obtain a stronger notion of convergence (cf.\ \cref{thm:sts}), we may place our analysis within the popular framework of graphon theory \cite{Lua21}, termed the \emph{graphon sampling model}.   

Recall that a graphon $\bW$ is a bounded symmetric measurable function
\begin{align*}
    \bW: [0,1]^2 \to [0,1],
\end{align*}
modeling an ``infinite'' graph. Intuitively, $\bW$ encodes edge connection probability. It induces a compact self-adjoint operator $T_{\bW}$ on $L^2([0,1])$, $(T_{\bW}h)(u) = \int_0^1 \bW(u,v)h(v)\ud v$, whose spectrum is countable, real, and accumulates only at $0$.

Given such a graphon $\bW$, we sample a nested sequence of graphs as follows. Let $(s_i)_{i\geq 1}$ be i.i.d.\ uniform samples from $[0,1]$, and for each pair $i<j$, independently, connect $v_i$ and $v_j$ by an edge with probability $\bW(s_i,s_j)$. The graph $G_n$ is the induced graph on $V_n = \{v_1,\ldots,v_n\}$. Such graphs are known as \emph{$\bW$-random graphs}, and they converge to $\bW$ (in the homomorphism density sense) with probability $1$ \cite{Lov:12, Lua21}.

A graphon signal is a function $x:[0,1]\to\mathbb{R}$. It induces the sampled vector $\bx_n$ with $\bx_n(i)=x(s_i)$. For continuous $x$, the strong law gives $n^{-1}\norm{\bx_n}_2^2\to\norm{x}_{L^2}^2$. We therefore set $\gamma_n=\sqrt{n}$ in \cref{eq:fmm} for \cref{thm:sts}, matching the normalization relating sampled GFT coefficients to graphon Fourier coefficients \cite[Lemma~2]{Lua21}.

The following sufficient condition allows each fixed low-index coefficient to be matched to an isolated negative graphon eigenmode. It is not necessary for convergence in general: for a constant graphon, normalized coefficients can instead converge to zero.

\begin{Assumption}[Negative spectral separation] \label{ass:spec}
The graphon operator $T_{\bW}$ has infinitely many negative eigenvalues, each of which is simple.
\end{Assumption}

Since the eigenvalues of \(\bS_n=\bA_n\) are ordered increasingly, the convergence established in \cref{thm:sts} below concerns the negative spectral branch of the graphon operator. The positive branch can be treated analogously by reversing the eigenvalue ordering. In particular, if \(T_\bW\) has infinitely many positive eigenvalues, each simple, then the same argument applied to the decreasingly ordered eigenvalues yields convergence to the corresponding positive graphon Fourier coefficients. We focus on the negative branch for clarity.

For ease of reference, we summarize the differences between the settings of the two graph sequence models in \cref{tab:cbt}. 
\begin{table}[!htb]
\renewcommand{\arraystretch}{1.5} 
    \caption{Comparisons between the graph sequence models.}
    \label{tab:cbt}
    \centering
    \begin{tabular}{|c|c|c|}
    \hline
       Model  & Random expansion & Graphon 
       sampling \\ 
       \hline
        Ambient structure & Countable $\widehat{G}$ & Graphon $\bW$ \\
         \hline
        Signal 
        & Unconditional $\bx \in \mathbb{R}^{\infty}$ & Continuous $x$ \\
         \hline
         Operator $\bS_n$ & Arbitrary & $\bA_n$ \\
         \hline
        Scaling $\gamma_n$ & Depends on $\bx_n$ & $\sqrt{n}$ \\
         \hline
         Spectral assumption & None & Assumption 1\\
         \hline
    \end{tabular}
\end{table}

\subsection{The convergence theorems} \label{sec:tct}

In this subsection, we present the convergence results anticipated in \cref{sec:tir}, which support the asymptotic consistency of the (absolute) spectral series. We first establish a uniform subsequential convergence result. We emphasize that it holds for any sequence of graphs, shift operators, and signals satisfying the stated norm bound, in particular, for every realization of the random expansion model in \cref{tab:cbt}; the graph structure enters through the GFT. It should therefore be best read as an asymptotic well-posedness property of the spectral-series construction. 

\begin{Theorem} \label{thm:tso}
Assume the set of partitions $\mathcal{P}$ is non-vanishing and has diminishing tails, the interpolation is regular, and $\norm{\bx_n}_2=O(\gamma_n)$. Then for every $t \in (0,1)$, the sequence $(f_n)_{n\geq 2}$ has a subsequence converging uniformly on $[0,t]$ to a function in $BV([0,t])$.
\end{Theorem}

For the graphon sampling model in the second half of \cref{sec:taf} d), we have a stronger uniform convergence, with an explicitly identified limit. Notice that, however, we do not claim that \cref{thm:sts} subsumes \cref{thm:tso}, as the results concern different scenarios.

\begin{Theorem} \label{thm:sts}
Consider the graphon sampling model with $\bS_n=\bA_n$, a continuous signal $x$, $\gamma_n=\sqrt{n}$, and \cref{ass:spec} (cf.\ \cref{tab:cbt}).\footnote{More generally, the same conclusion holds for any fixed measurable $x\in L^2([0,1])$ by a density argument given in Appendix~\ref{app:pot}.} Suppose $\mathcal{P}$ is non-vanishing, index-wise convergent with limits $a_i=\lim_{n\to\infty}|I_{i,n}|>0$, and has diminishing tails. If the interpolation is regular and consistent, then almost surely (a.s.) $f_n$ converges uniformly on every $[0,t]$, $0<t<1$, to a function $f\in BV([0,t])$.
\end{Theorem}

For every fixed $t<1$ and truncation level $\eta$, these results also give convergence of the signatures of the normalized, time-augmented paths \emph{restricted to $[0,t]$}:
\begin{equation*}
S^{\leq\eta}_{(f_n,\mathrm{id})}([0,t])\longrightarrow
S^{\leq\eta}_{(f,\mathrm{id})}([0,t]).
\end{equation*}
For \cref{thm:tso} the convergence is along the selected subsequence; for \cref{thm:sts} it holds almost surely along the full sequence. Indeed, uniform convergence and the uniform variation bounds in Appendix~B imply convergence in $p$-variation for $1<p<2$, where truncated signatures are continuous \cite{Fri12, Lyo14}. In practice, this allows models to effectively learn and estimate from localized (small) subgraphs.

\section{Numerical results} \label{sec:nr}
In this section, we present numerical results to demonstrate the theoretical insights and practical utility of our model. We depart from the strict asymptotic setup of \cref{sec:cog} (in our finite regime) to utilize conventional configurations of graph Laplacian for GSO, uniform partition, and linear interpolation.

\subsection{Trend prediction: Covid-19 dataset} \label{sec:tpc}
We consider a Covid-19 dataset from The New York Times.\footnote{\url{https://github.com/nytimes/covid-19-data}} We use records for $n=58$ counties in California from 2020 to 2022. The dataset records the daily infection count for each county. We consider the \emph{trend prediction} problem: given a period of consecutive observations, we want to make a binary decision to predict whether, on average, the number of infections will either \emph{increase or decrease} after $3$ days. 

\begin{figure}[!htb]
    \centering
    \includegraphics[width=0.8\linewidth]{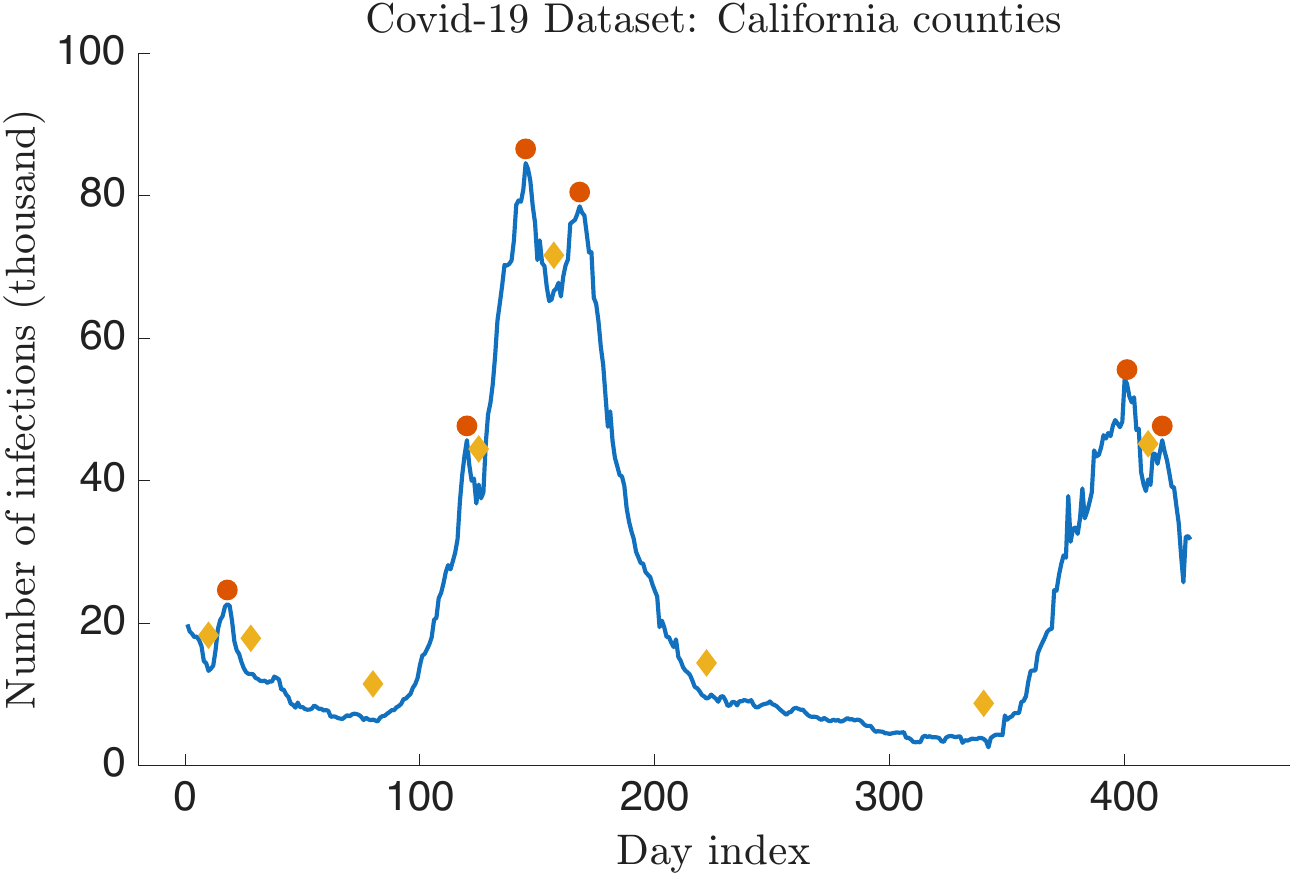}
    \caption{We show the average daily infection count over all counties. Local maxima and minima are highlighted. }
    \label{fig:wst}
\end{figure}

In \cref{fig:wst}, we show the curve of the average number of infections, with local maxima and local minima highlighted. Notice that small random fluctuations are disregarded. For the ground truth, we consider the curve to have a decreasing (resp. increasing) trend while it moves towards a local minimum (resp. maximum). Such a trend is the average observation over the entire state, as a local maximum or minimum usually arrives on different dates for different counties.  

To proceed, we first construct the time-vertex graph $G$. Specifically, let $H_0$ be the graph whose nodes correspond to counties and whose edges are defined by geographic adjacency. On the other hand, let $H_1$ be the path graph with $T$ nodes for the time domain. Then $G = H_0\square H_1$ is the Cartesian product. Its Laplacian matrix is related to those of $H_0, H_1$ via the formula $\bL_{G} = \bL_{H_0}\otimes \bI_T + \bI_{58}\otimes \bL_{H_1}$. 

A node of $G$ is a pair $(v,t)$, where $v$ corresponds to a county and $t$ is a day (index). A signal on $(v,t)$ is the number of infections for the county $v$ on day $t$. With this interpretation, we obtain a signal $\bx_{t_0}$ on $G$ for the $T$-day period starting from $t_0$. An example of the spectrum of such a graph signal is shown in \cref{fig:wse}. It displays a non-bandlimited behavior, which our signature approach is designed to address. Its label $y_{t_0}=1$ or $0$ is the trend, either increasing or decreasing, on day $t_0+T+2$.  

For the experiment, we choose $T=7$ and construct a dataset of graph signals as above, such that the training/validation/test ratio is approximately 15/1/1. To ensure no data leakage, there is no time overlap between any pair of training and test signals.  

To apply the signature method, given a signal and its label $(\bx_{t_0},y_{t_0})$ as above, we fix the maximum index length $\eta>0$, and construct $\bs_{\bx_{t_0},\eta}$ defined in \cref{defn:tsb}. Here, the standard linear interpolation is used. We then apply a trainable classifier $C_{\theta}(\bs_{\bx_{t_0},\eta})$ to predict its label $y_{t_0}$. We consider different choices of $C_{\theta}(\cdot)$: LinearSVM, LogisticElasticNet, MLP; LightGBM, RandomForest, XGBoost \cite{Mou25}. 

The study has the following \emph{objectives}: (a) to examine how signature length affects model performance; (b) to provide empirical evidence for \cref{thm:sti}, which suggests signature features couple well with simple linear classifiers such as LinearSVM; (c) to understand model performance for an expansion of subgraphs. 

\begin{figure}[!htb]
    \centering
    \includegraphics[width=0.8\linewidth]{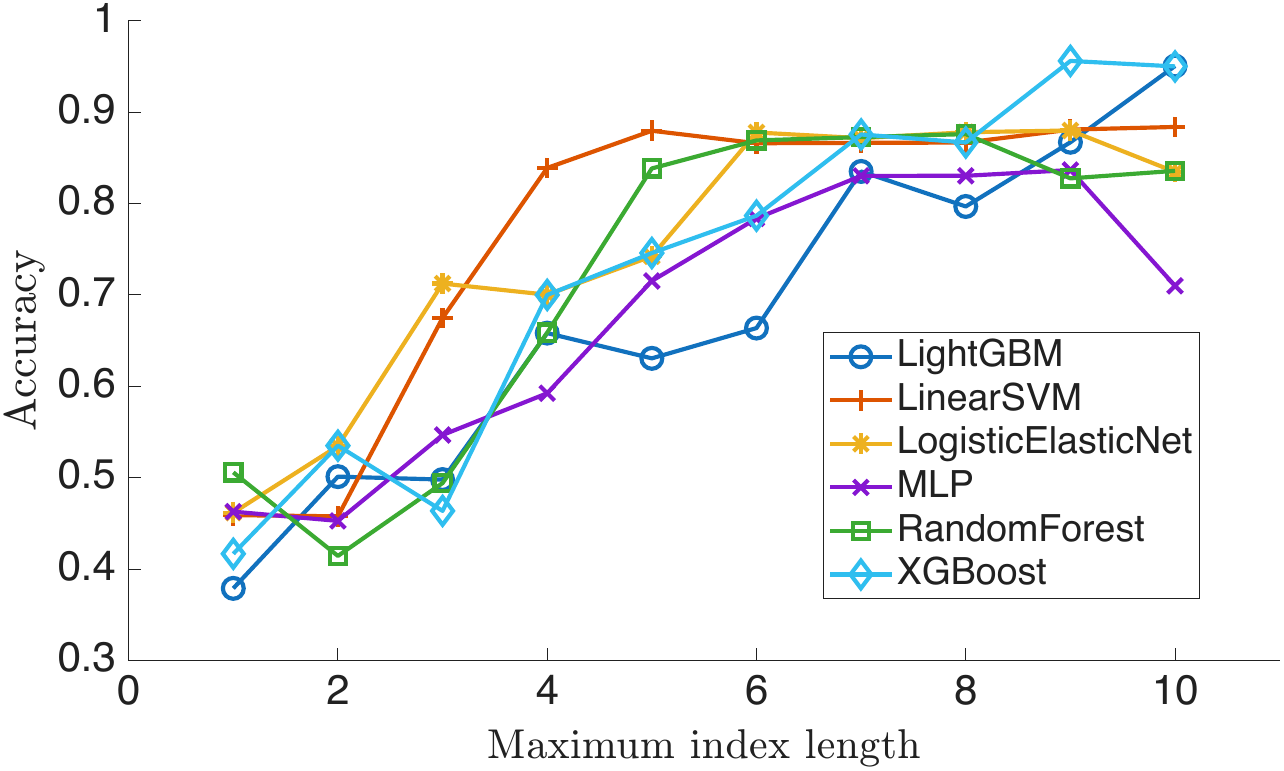}
    \caption{We show the results for different classifiers. We observe that in general, the performance improves with the maximum length of the signature index.}
    \label{fig:fdc}
\end{figure}

We show the test accuracy in \cref{fig:fdc}. A few observations are in order. In general, the accuracy indeed increases w.r.t.\ $\eta$, regardless of the classifier used. For individual classifiers, both LightGBM and XGBoost start with low accuracies and gradually achieve a high accuracy of $\approx 96\%$ at $\eta=9, 10$. This suggests that the signature features $\bs_{\bx_{t_0},10}$ are sufficiently expressive for the prediction task. For lighter linear classifiers, a small $\eta \approx 5$ achieves reasonably good performance, and increasing $\eta$ provides limited gains. This observation suggests that more important global features are sufficiently encoded for small $\eta$.  

In view of \cref{thm:sti}, we also compare predictive performance with classifier complexity. For the linear classifiers LinearSVM and LogisticElasticNet, and the non-linear MLP, we show their total number of parameters in \cref{tab:mcm}. For the remaining tree-based models, we show the total number of decision tree leaves in \cref{tab:mcm}. 

It is interesting to notice that linear models LinearSVM and LogisticElasticNet both outperform MLP, though the latter has substantially more parameters. For example, LinearSVM exceeds $85\%$ accuracy ($\eta=5$) with only $63$ parameters. This aligns with the universal linear approximation \cref{thm:sti}. For tree-based models, it remains true that XGBoost with fewer parameters generally performs better.

\begin{table}[!htb]
\renewcommand{\arraystretch}{1.5} 
    \caption{Model complexity measured in terms of the total number of parameters or the total number of leaves, used for the top performance in \cref{fig:fdc}. (LEN: LogisticElasticNet.)}
    \label{tab:mcm}
    \centering
    \begin{tabular}{cccccc}
    \hline
    LinearSVM & $63$ &
    LEN  & $127$ &
    MLP & $40961$ \\
    LightGBM  & $5504$ &
    RandomForest & $10218$ & 
    XGBoost & $2600$ \\
         \hline
    \end{tabular}
\end{table}

\begin{figure}
    \centering
    \includegraphics[width=0.8\linewidth]{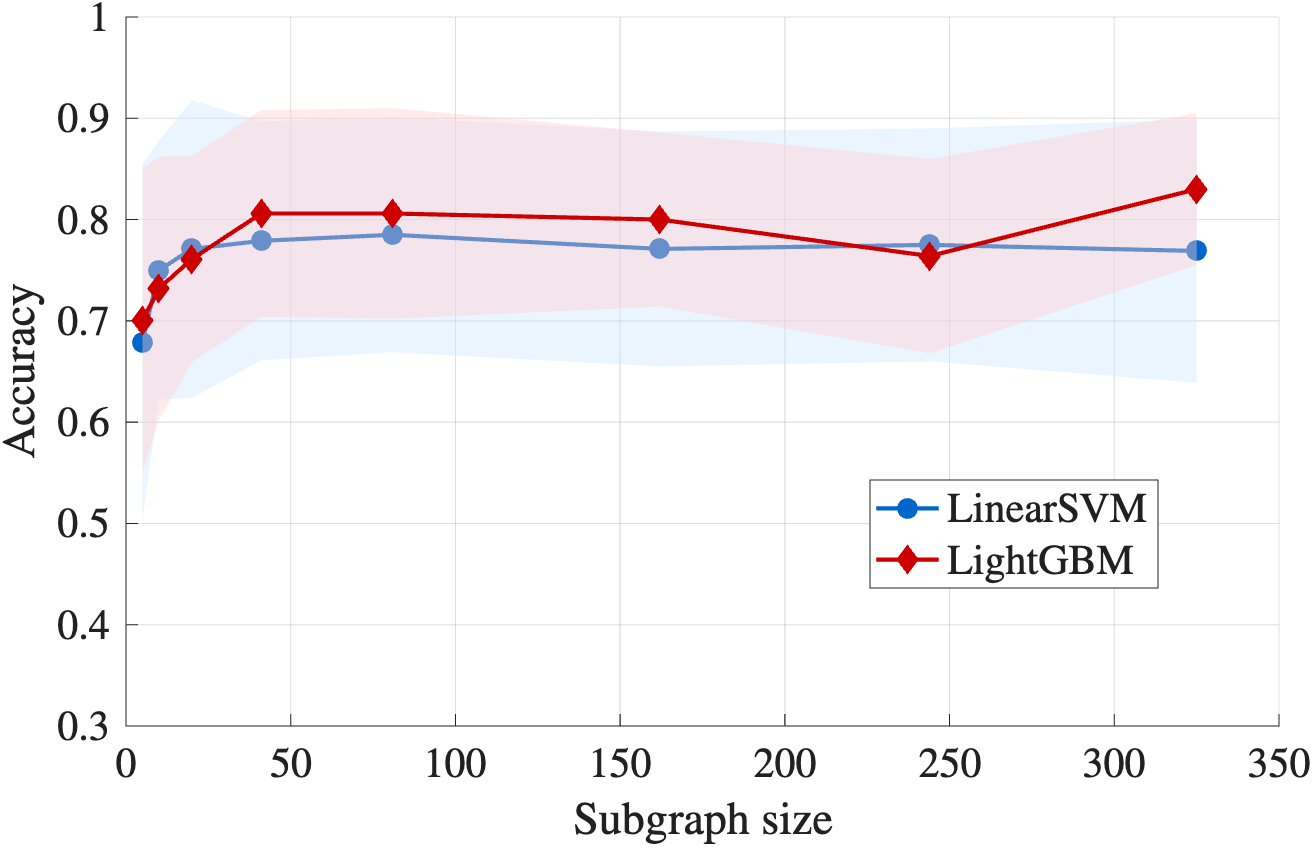}
    \caption{The prediction accuracy (mean and standard deviation) when LinearSVM or LightGBM are applied to a sequence of expanding subgraphs.}
    \label{fig:tpc}
\end{figure}

We further consider an expansion of subgraphs (cf.\ \cref{sec:cog}). Specifically, starting from a random seed node of $G$, we generate a sequence of subgraphs and apply the proposed approach only using the subgraph structures and the signals on them. Notice that in these cases, the exact physical meaning, location, or date of each node is unknown. We run LinearSVM ($\eta=5$) and LightGBM ($\eta=10$), which are the top performers on the full graph. The results on $30$ random seeds are shown in \cref{fig:tpc}. We see that the performance for both approaches quickly stabilizes, which is faster for the lighter linear model. This gives empirical evidence for the discussion at the end of \cref{sec:cog}.

\subsection{Change detection: the ECoG dataset}
We conduct change point detection (CPD) experiments on a brain ECoG dataset,\footnote{\url{https://math.bu.edu/people/kolaczyk/datasets.html}} consisting of normalized time series recordings from $76$ brain electrodes over $10$ seconds each.  The signals have a sample rate of $ 400$ Hz and are collected from $5$ epilepsy patients (denoted by pt 1 to pt 5), during two seizure-related periods: pre-ictal (normal) and ictal (abnormal).

Let $H_0$ be the electrode graph constructed as in \cite{Kra08}, and $H_1$ be the (temporal) path graph with $10$ nodes. As in \cref{sec:tpc}, we construct a spatio-temporal graph $G = H_0\square H_1$, and $\bL_{G}=\bL_{H_0}\otimes\bI_{10}+\bI_{76}\otimes\bL_{H_1}$. Therefore, a signal $\bx$ on $G$ represents readings of all electrodes every $ 25$ ms.   

We consider both online and offline CPD as follows:
\begin{itemize}
    \item OCPD: We consider an online streaming of pre-ictal signals on $G$ for a patient. At a uniformly random discrete time-stamp $t^*$ in $[100, 150]$, the signal streaming becomes ictal. One is required to identify the change as soon as possible during streaming.\footnote{To avoid a trivial solution, we demean the signal at each node of $G$.}  
    \item CPD: We make a full observation of a time series of signals on $G$. Assume that there are $2$ randomly chosen change points $t_1^* < t_2^*$ from pre-ictal condition to ictal condition and then back to pre-ictal condition. We want to estimate both $t_1^*$ and $t_2^*$ simultaneously.   
\end{itemize}

Our approach essentially produces signature features $\bs_{\bx,\eta}$ from input signals. Motivated by the previous study, we choose a small $\eta=3$, which makes OCPD fast. To use these features, we need a base multivariate change point detection model, such as a kernel-based model. For OCPD (resp.\ CPD), we consider KCUSUM \cite{Fly19} (resp. KCP \cite{Arl19}) as the base model.  

Both KCUSUM and KCP require a kernel function of features. We consider the following choices:
\begin{itemize}
    \item The linear kernel: $k(\bx,\by) = \langle \bx,\by \rangle$.

\item The Gaussian kernel: $k(\bx,\by) = \exp(-\norm{\bx-\by}^2/\sigma)$.
\item The Laplace kernel: $k(\bx,\by) = \exp(-\norm{\bx-\by}/\sigma)$.
\end{itemize}
These kernel functions can be applied to both the original graph signals $\bx$ and their converted (truncated) signatures $\bs_{\bx,\eta}$. 

The primary \emph{objective} of this study is to determine whether converting a graph signal into its signatures is beneficial for the detection tasks. For more baselines that handle non-bandlimited signals, we consider applying graph wavelet \cite{Ham11} and scattering transforms \cite{Zou20, Zou25} to the graph signals, whose principal components are used for KCUSUM and KCP.

To evaluate, we report the percentage of cases $r \in [0,1]$ where either kernel method successfully detects the change point(s). For OCPD, we monitor the detection delay $\tau$ averaged over multiple runs per patient, and a false alarm is considered a failure. On the other hand, for CPD, we interpret the task as a segmentation problem, and if $\overline{t_1}, \overline{t_2}$ are the estimated change points, we report the estimation error $\epsilon_1 = |\overline{t_1}-t_1^*|$ and $\epsilon_2 = |\overline{t_2}-t_2^*|$, averaged over multiple runs. 

\begin{table}[!htb]
\centering
\caption{OCPD: the average detection delay $\tau$ for different patients and kernel functions. In (brackets), we show the successful detection rate $r$. The best result in each column is {\bf boldfaced}.} \label{tab:ota}
\centering
\begin{tabular}{cccccc} 
\toprule
Kernel & pt 1 &pt 2 &pt 3 &pt 4 &pt 5 \\ 
\midrule
Linear& {\bf 9.4} (1.0)& {\bf 17.8} (1.0)& {\bf 15.8} (1.0)& {\bf 10.4} (1.0)& 19.6 (1.0)\\ 
Gauss.& 16.6 (1.0)& 59.3 (0.8)& 26.2 (1.0)& 35.6 (1.0)& {\bf 6.8} (1.0)\\ 
Laplace & 62.2 (1.0) & 69.3 (0.6)& 21.0 (0.6)& 52.4 (1.0)& 19.2 (1.0)\\ 
\bottomrule
\end{tabular}
\end{table}

\begin{table}[!htb]
\centering
\caption{OCPD: we show the best performance among the three baselines.} \label{tab:wst}
\centering
\begin{tabular}{cccccc} 
\toprule
Kernel & pt 1 &pt 2 &pt 3 &pt 4 &pt 5 \\ 
\midrule
Linear& -- (0.0)& 9.0 (0.2)& 42.3 (0.6)& 30.8 (0.8)& 26.8 (1.0)\\ 
Gauss.& -- (0.0)& 10.0 (0.2)& 30.0 (0.6)& 38.7 (0.6)& 45.8 (1.0)\\ 
Laplace & -- (0.0) & 9.0 (0.2)& 44.0 (0.8)& 16.0 (0.2)& 48.6 (1.0)\\ 
\bottomrule
\end{tabular}
\end{table}

\begin{figure}[!htb]
    \centering
    \includegraphics[width=.48\linewidth]{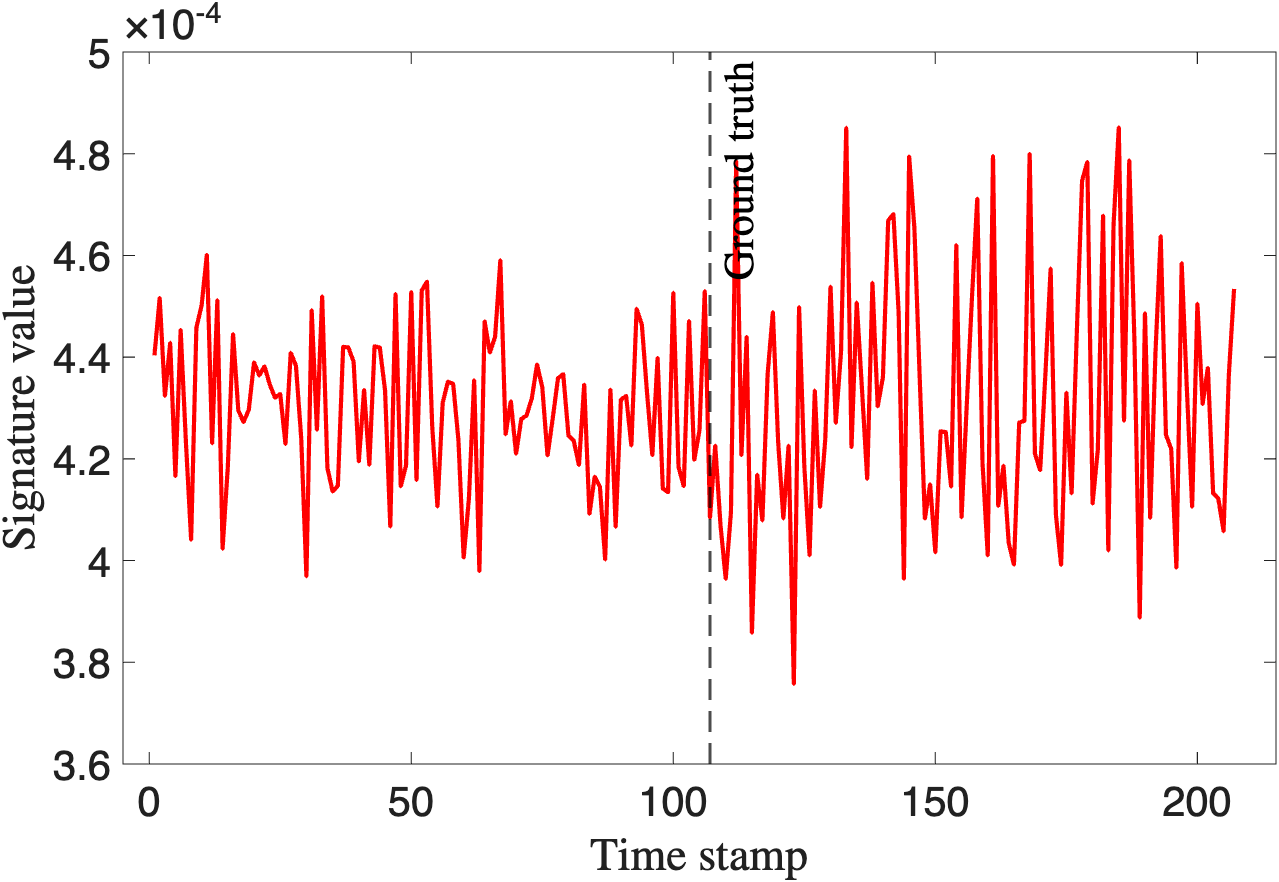}\hfill
    \includegraphics[width=.48\linewidth]{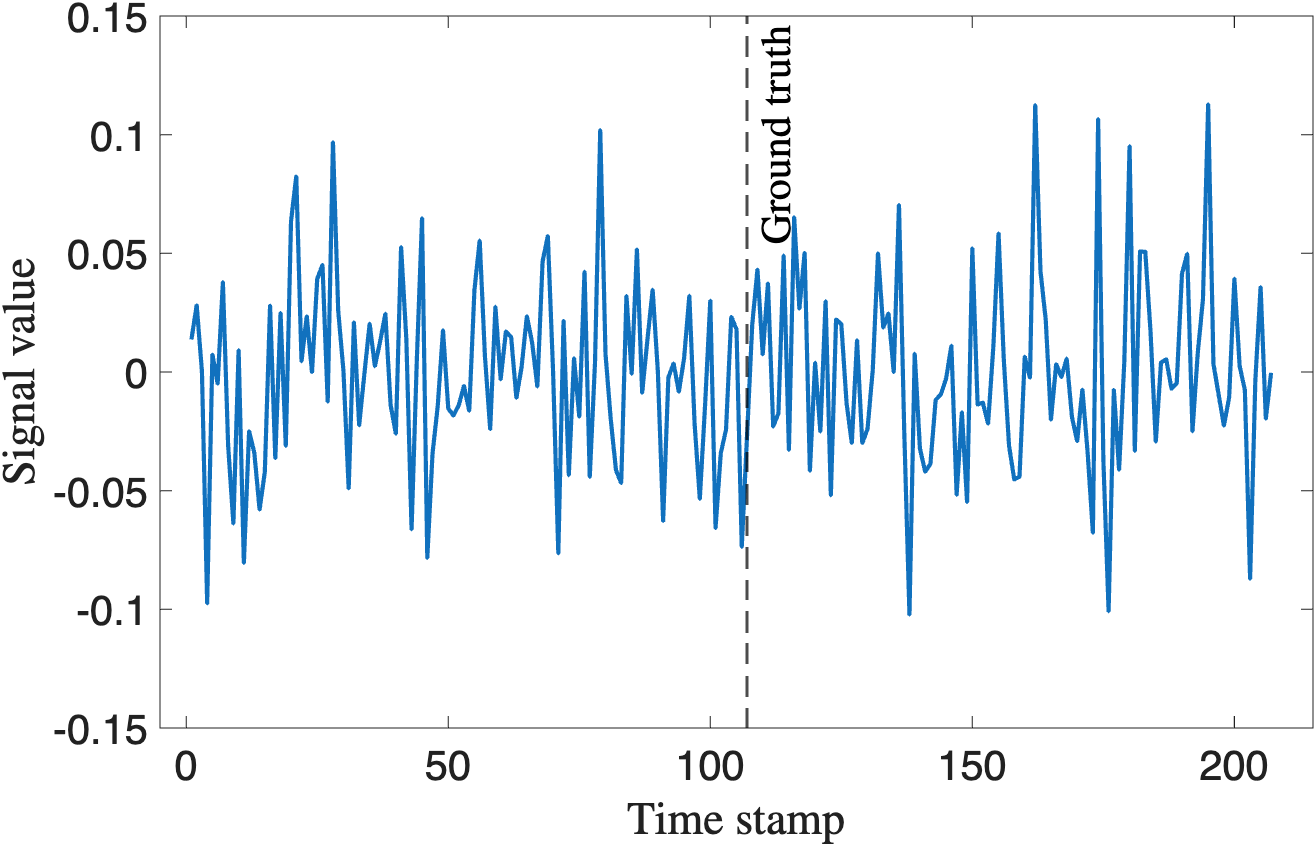}
    
    \vspace{0.5em} 

    \includegraphics[width=.48\linewidth]{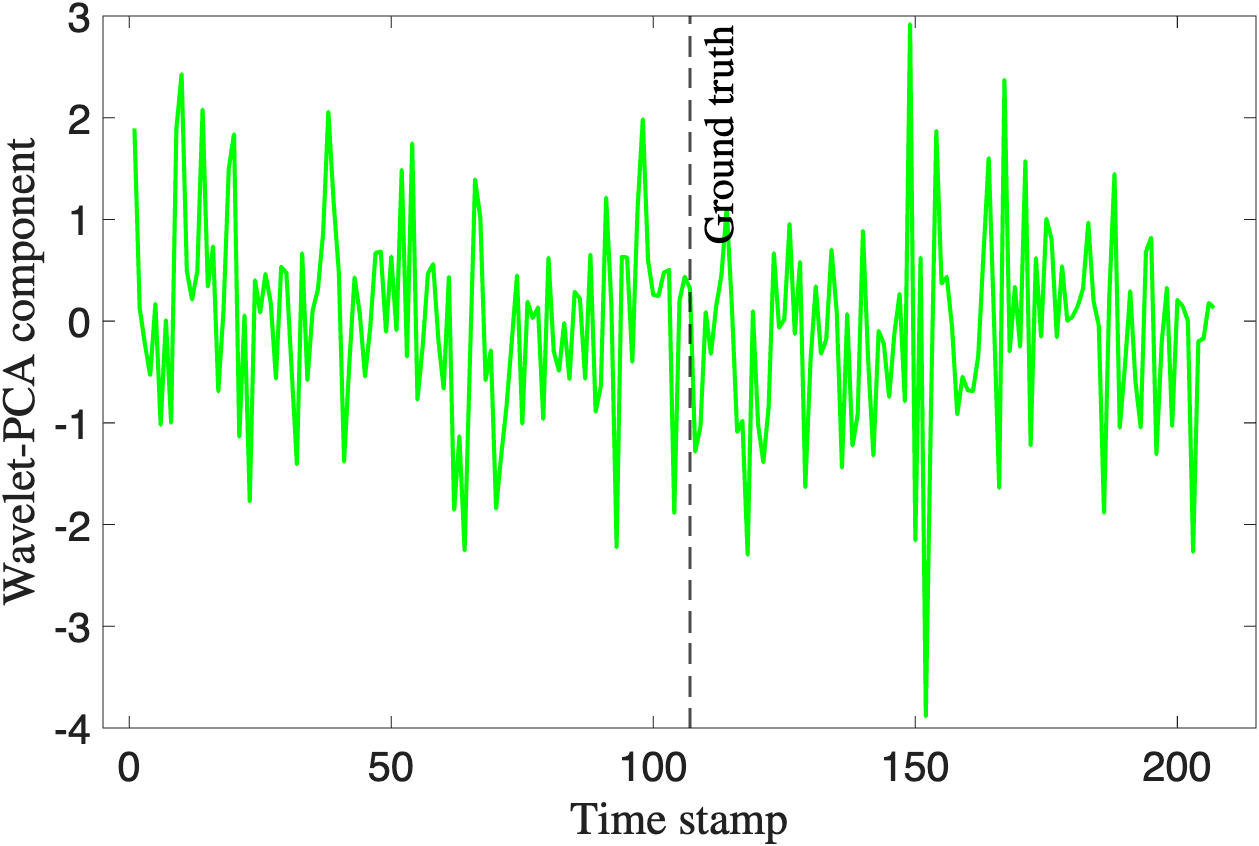}\hfill
    \includegraphics[width=.48\linewidth]{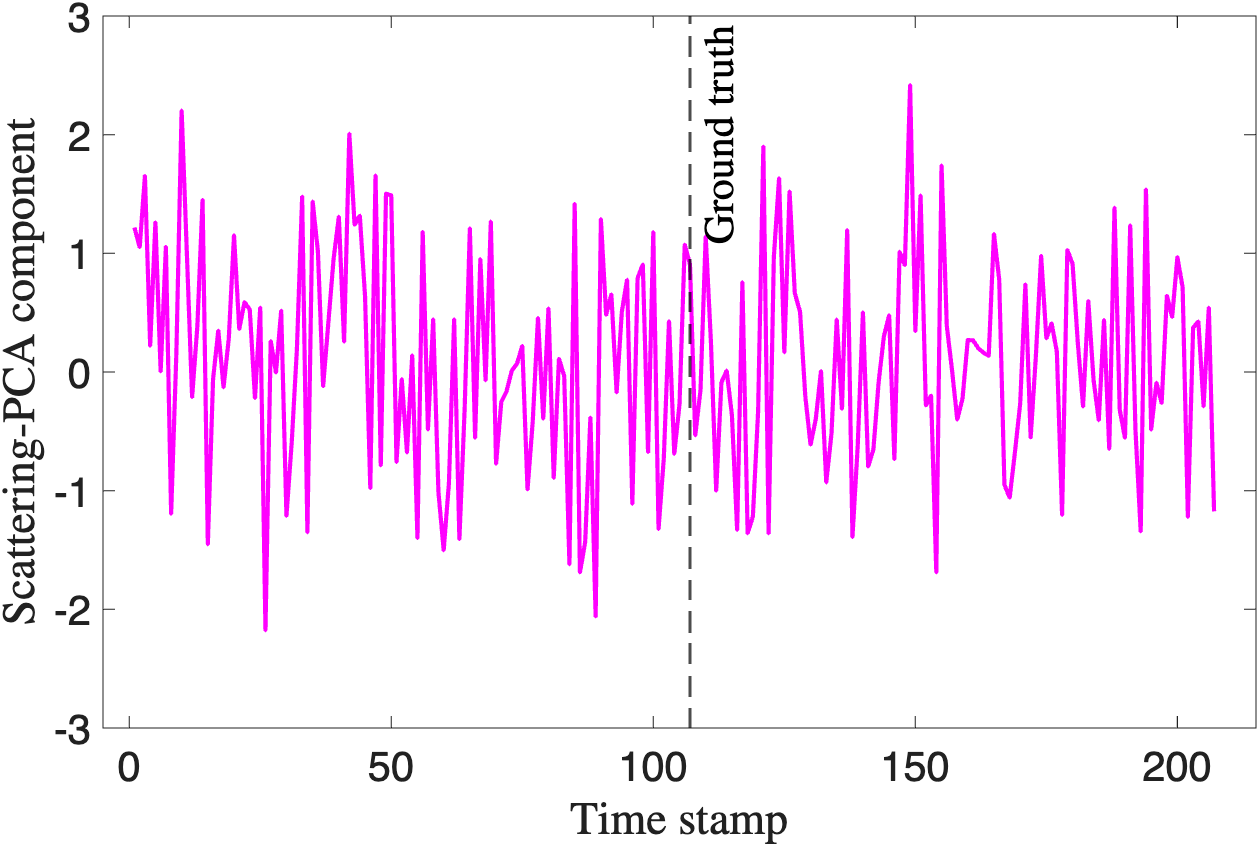}
    
    \caption{For illustration with pt 1, in the $4$ panels, we show the time series of a random component from the (feature) vector obtained via the signature method, raw signal, wavelet, and scattering transforms. The signature panel shows a more prominent pre- and post-change distribution difference.}
    \label{fig:itt}
\end{figure}

The results for OCPD are shown in \cref{tab:ota}. We notice that with converted signature features, KCUSUM makes successful detections in most cases, and the linear kernel suits the task best with a generally low detection delay. For \emph{comparison}, we show the entry-wise best performance among the $3$ baselines in \cref{tab:wst}. Overall, the signature method outperforms the baselines. This can be attributed to a more prominent distribution shift of (transformed) signals after the change point, as illustrated in \cref{fig:itt}.  

Owing to the small truncation length $\eta$, the average runtime per online detection is \emph{only $1.47$ ms} across all patients, making the approach highly viable for time-critical OCPD applications.

The results for KCP-based CPD are shown in \cref{tab:cts} and \cref{tab:cta}. Consistent with KCUSUM, we see that the baselines have very low detection rates (and hence we do not show their estimation errors). Using signature features, both the Laplace kernel and the Gaussian kernel give reasonably good results. We observe that there are cases with large $\epsilon_2$. This is mainly due to the fact that for one or two outliers in our samples, KCP performs poorly with $\epsilon_2 > 150$. However, in general, the method is strong for detecting the first change point from pre-ictal to ictal conditions, which can be more relevant in practice.
\begin{table}[!htb]
\centering
\caption{CPD: The successful detection rate $r$, and in (brackets), we show the best rate among the baselines.} \label{tab:cts}
\centering
\begin{tabular}{cccccc} 
\toprule
Kernel & pt 1 &pt 2 &pt 3 &pt 4 &pt 5 \\ 
\midrule
Linear& 0.8 (0.2) & 1.0 (0.2) & 0.8 (0.0)& 0.8 (0.0)& 0.6 (0.2)\\ 
Gauss.& 1.0 (0.0) & 0.8 (0.2) & 0.8 (0.0)& 0.8 (0.2) & 1.0 (0.4)\\ 
Laplace & 1.0 (0.2) & 1.0 (0.2) & 0.8 (0.2)& 0.8 (0.2) & 1.0 (0.2)\\ 
\bottomrule
\end{tabular}
\end{table}

\begin{table}[!htb]
\centering
\caption{CPD: The average estimation errors $\epsilon_1, \epsilon_2$. The average is taken over successful two change point detections for the signature method.} \label{tab:cta}
\centering
\begin{tabular}{cccccc} 
\toprule
Kernel & pt 1 &pt 2 &pt 3 &pt 4 &pt 5 \\ 
\midrule
Linear& 165, 128& 147, 386 & 79.8, 281& 72.0, 171 & 79.0, 207 \\ 
Gauss.& 11.6, 2.0 & 0.5, 4.8 & 9.5, 2.3 & 3.8, 3.5 & 4.8, 44.8\\ 
Laplace & 0.8, 1.6 & 1.0, 48.8 & 0.0, 1.3 & 2.3, 1.3 & 3.0, 43.9 \\ 
\bottomrule
\end{tabular}
\end{table}

\paragraph*{Summary} From both studies in \cref{sec:nr}, we notice reasonable performance for small $\eta$. This aligns with the philosophy of traditional GSP for bandlimited graph signals, that is, to use a low-dimensional vector in signal representations. For non-bandlimited signals, our signature approach (partially) achieves a similar objective, as small-length signatures encode most useful global characterizations of the signals. 

\section{Conclusions} \label{sec:con}
This paper introduces a novel signature-based framework for analyzing graph signals that are not well described by bandlimited models. The framework interprets an ordered graph Fourier spectrum as a graph spectral series and uses path signatures to summarize its global spectral structure. We establish theoretical properties that support this construction and provide conditions under which graph spectral series admit an ordered-path interpretation. Experiments on real-world trend-prediction and change-detection tasks demonstrate the practical value of the resulting features. Together, these results position graph spectral series as a principled representation for analyzing non-bandlimited graph signals.

\paragraph*{Future work} From a broader perspective, we identify promising directions for immediate future research:
\begin{itemize}
    \item This work establishes an initial connection between GSP and signature methods. To keep the discussion focused, we deliberately restrict ourselves to the foundational aspects of signature theory. Fully exploiting its power \cite{Fri12, Fra19, Tam25} requires considerable further development and is essential for additional applications.
    \item Incorporating the explicit values of the graph shift operator's eigenvalues into the spectral series construction could yield richer geometric representations. Concurrently, theoretically and empirically characterizing the optimal sorting or permutation ordering of GFT components presents an open challenge.
    \item Developing robust inverse algorithms to reconstruct the original graph signals from their signature features, analogous to sampling and recovery theorems for bandlimited signals \cite{Ani16, Cha18, She24}, carries significant practical and theoretical importance.
    \item While convergence results successfully demonstrate that the spectral-series construction is asymptotically well-posed, they represent an initial step with clear avenues for growth. Future work can extend the current framework to fully capture the high-index spectral tail near $1$. Additionally, we hope to investigate unique limits and explore ways to relax the spectral-separation assumptions.
\end{itemize}
Ultimately, we hope this framework catalyzes a deeper, systematic integration of signature theory into graph signal processing.

\appendices

\section{Proofs of signature properties}
For the benefit of readers unfamiliar with signature theory, we make all proofs as self-contained as possible.

\begin{IEEEproof}[Proof of \cref{thm:sbi}]
(a) Since the paths include the identity coordinate, equality of signatures implies equality of the remaining coordinates up to translation \cite[Proposition 2.34]{Che26}. Thus, if $\bs_\bx=\bs_{\bx'}$, evaluation at the partition endpoints gives $\widehat{\bx'}=\widehat{\bx}+c\mathbf{1}$ for some $c\in\mathbb{R}$, i.e., $\widehat{\bx'}$ most be on a fixed line $\ell$ in $\mathbb{R}^n$. For unit-norm signals, both coefficient vectors lie on $\mathbb{S}^{n-1}$ because the GFT is norm-preserving. The line $\ell$ meets the sphere $\mathbb{S}^{n-1}$ in at most two points, so there is at most one coefficient vector other than $\widehat{\bx}$ with the same signature. Invertibility of the GFT proves (a).

(b) Fix an orthogonal GFT matrix $\bU$, and let $\bU'$ be any orthogonal GFT matrix yielding the same signature of $\bX$. The same argument gives
\begin{align*}
\bU'^\top\bx_i=\bU^\top\bx_i+c_i\mathbf{1},\qquad 1\leq i\leq n.
\end{align*}
Both transformed vectors have the same norm $\norm{\bx_i}$. Hence, as in (a), for each $i$, there are at most two possibilities for $\bU'^\top\bx_i$, and only finitely many possible ordered collections $(\bU'^\top\bx_i)_{i=1}^n$. Since the $\bx_i$ form a basis, each collection uniquely determines $\bU'$. This proves finiteness of the admissible GFT matrices.
\end{IEEEproof}

\begin{IEEEproof}[Proof of \cref{thm:sti}]
(a) Equip $BV([0,1])$ with the norm $\norm{f}_{BV}=|f(0)|+V(f;[0,1])$, and use the sum of these norms for vector-valued paths. The truncated signature is continuous in this topology \cite[Proposition 7.66]{Fri12}. It therefore suffices to prove continuity of $\iota:\bX\mapsto\boldf_{\bX}$ from \cref{eq:bbf}. 

Notice that $\iota$ is obtained from first performing GFT $\bx \mapsto \widehat{\bx}$ and then the interpolation $\widehat{\bx} \to p(\mathcal{P}_n; \widehat{\bx})$, to each signal in $\bX$. The GFT is linear, and the time coordinate is fixed, so we need only check continuity of $\ba\mapsto p(\mathcal{P}_n;\ba)$ for $\ba\in\mathbb{R}^n$.

For $\ba,\ba'\in\mathbb{R}^n$, write $g=p(\mathcal{P}_n;\ba)-p(\mathcal{P}_n;\ba')$, $z_i=(\ba_i,\ba_{i+1})$, and $z_i'=(\ba_i',\ba_{i+1}')$. Put $h_i=|I_{i,n}|$, $u_i(y)=(y-e_{i,n})/h_i$, and $D(z,u)=\partial_3\phi(z_1,z_2,u)$. The chain rule and the substitution $y=e_{i,n}+h_i u$ give
\begin{equation}\label{eq:vbb}
\begin{split}
V(g;I_{i,n})
&=\frac{1}{h_i}\int_{I_{i,n}}\big|D(z_i,u_i(y))\\
&\qquad-D(z_i',u_i(y))\big|\ud y\\
&=\int_0^1|D(z_i,u)-D(z_i',u)|\ud u.
\end{split}
\end{equation}
Fix $\ba$. For $\ba'$ in a compact neighborhood of $\ba$, all endpoint pairs lie in a common compact subset $K_0$ of $\mathbb{R}^2$. Strong consistency makes $D$ uniformly continuous on the product space $K_0\times [0,1]$. Hence each integral in \cref{eq:vbb} tends to zero as $\ba'\to \ba$, uniformly over the finitely many $i$. Additivity of then gives 
\begin{align*}
V(g;[0,1])=\sum_{1\leq i<n} V(g; I_{i,n}) \to 0.
\end{align*}
Moreover, $g(0)=\ba_1-\ba_1'\to0$, proving continuity in the full BV norm and thus continuity of $\sigma_\eta$.

(b) We have seen that GFT and the interpolation induce a function $\iota: \mathbb{R}^{n\times d} \to BV([0,1])^{d+1}, \bX \mapsto \boldf_{\bX}$ (cf.\ \cref{tab:nl}). It is injective as both GFT and the interpolation are injective. 

Write $\boldf_{\bX}=(f_1,\ldots,f_d,\mathrm{id})$ and define the based, time-augmented path
\begin{align*}
\widetilde{\iota}(\bX)
=(f_1-f_1(0),\ldots,f_d-f_d(0),\mathrm{id}).
\end{align*}
Notice that the following holds for $\widetilde{\iota}$:
\begin{itemize}[leftmargin=1.5em]
  \item The map $\widetilde{\iota}$ is continuous. Notice that $f_j(0)$ is the leading GFT coefficient of $\bx_j$, and it is clearly continuous in $\bX$. Therefore, the continuity of $\widetilde{\iota}$ follows from that of $\iota$ proved in (a). 
  \item The map $\widetilde{\iota}$ is injective on $K$. To see this, if $\widetilde{\iota}(\bX) = \widetilde{\iota}(\bX')$, then for each $j\leq d$, $f_j=f_j'+c_j$ for some $c_j\in \mathbb{R}$. Trivially, the same identity holds for the GFTs of $\bx_j$ and $\bx_j'$, whose interpolations are $f_j$ and $f_j'$, respectively. By the assumption on $K$, we have $\bX = \bX'$.  
  \item For any $\bX$ and truncation length $\eta$, $\iota(\bX)$ and $\widetilde{\iota}(\bX)$ have the same signature $\sigma_{\eta}(\bX)$. This follows that path integrals of functions are translation invariant.   
\end{itemize}

In summary, $\widetilde{\iota}$ is a continuous bijection from $K$ to its image. As $K$ is compact, it induces a homeomorphism from $K$ to $\widetilde{\iota}(K)$, and the latter is thus also compact. The continuous function $\gamma$ thus induces a continuous function $\gamma\circ \widetilde{\iota}^{-1}: \widetilde{\iota}(K) \to \mathbb{R}$. 

We apply \cite[Proposition 2.35]{Che26} to $\widetilde{\iota}(K)$ to conclude that: for any $\epsilon>0$, there is $\eta$ and $\ell$ such that for any $\bX\in K$, we have 
\begin{align*}
\left|\gamma\circ \widetilde{\iota}^{-1}\Big(\widetilde{\iota}(\bX)\Big) - \ell\circ\sigma_{\eta}(\bX)  \right| < \epsilon, 
\end{align*}
which proves (b).
\end{IEEEproof}

\section{Proofs of the convergence results} \label{app:pot}
\begin{IEEEproof}[Proof of \cref{thm:tso}]
Step~1. Fix $t\in(0,1)$ and consider $n\geq2$. Diminishing tails gives integers $N\geq2$ and $M_0>N$ such that
\begin{align*}
\sum_{i=N}^{n-1}|I_{i,n}|<1-t \qquad(n\geq M_0).
\end{align*}
Thus the first $N-1$ intervals cover $[0,t]$ for $n\geq M_0$. Taking $K_t=\max(N,M_0)$ also covers the finitely many smaller $n$, since all their intervals are included. Consequently,
\begin{align*}
[0,t]\subseteq\bigcup_{i=1}^{\min(K_t,n-1)}I_{i,n}
\qquad(n\geq2).
\end{align*}
For each of the finitely many indices $i\leq K_t$, non-vanishing bounds $|I_{i,n}|$ away from zero for all sufficiently large $n$. The remaining, finitely many defined intervals have positive lengths. Hence
\begin{align*}
a:=\inf_{\substack{n\geq2\\1\leq i\leq\min(K_t,n-1)}}|I_{i,n}|>0.
\end{align*}
These bounds hold for the full sequence; no subsequence of interval lengths is required.

Step~2. We next show that when restricted to $[0,t]$, the sequence of functions $(f_n)_{n\geq 2}$ is uniformly bounded. 

We first notice that $f_n$ when restricted to the endpoints of the intervals of $\{I_{i,n} \mid 1\leq i\leq \min(K_t,n-1)\}$ is uniformly bounded, i.e., there is an upper bound $M$ for $|\widehat{\bx_n}(i)/\gamma_n|$ that is independent of $i$ and $n$. To see this, let $u_{i,n}$ be $i$-th eigenvector of $S_n$. Then, by the Parseval identity (for GFT), we have
    \begin{align*}
        |\widehat{\bx_n}(i)/\gamma_n| \leq \norm{\widehat{\bx_n}/\gamma_n} = \norm{\bx_n/\gamma_n}_2 = O(1), 
    \end{align*}
    which is the claim.
    
For a general $y\in[0,t]$ and any $n\geq 2$, there is an index $1\leq i\leq\min(K_t,n-1)$ such that $y\in I_{i,n}$, with endpoints $e_{i,n}$ and $e_{i+1,n}$. As the interpolation function is regular, we have 
        \begin{align*}
        &|f_n(y)-f_n(e_{i,n})| \\ \leq &\frac{g(|f_n(e_{i+1,n})-f_n(e_{i,n})|)}{e_{i+1,n}-e_{i,n}}\big(y-e_{i,n}\big)\\ \leq &g(2M). 
        \end{align*}
Hence, $|f_n(y)|\leq g(2M)+M$, independent of $n$. 

Step~3. We claim that $(f_n)_{n\geq 2}$ is uniformly equicontinuous when restricted to $[0,t]$. Set $B=g(2M)/a$. By regularity, we notice that on each relevant interval $I_{i,n}$, the Lipschitz constant of $f_n$ is at most $B$. Moreover, consider $0\leq y < z\leq t$ in different intervals such that $y\in I_{i,n}$ and $z\in I_{j,n}$, $i< j$. Split $[y,z]$ at the partition endpoints
\begin{align*}
    [y,z] = [y,e_{i+1,n}]\cup I_{i+1,n}\cup \ldots I_{j-1,n}\cup [e_{j,n},z].
\end{align*}
Add the resulting bounds to obtain
\begin{align*}
&|f_n(z)-f_n(y)|\\ \leq & \,B\!\left(e_{i+1,n}-y +\sum_{k=i+1}^{j-1}|I_{k,n}| +z-e_{j,n}\right) \\
=\, &B(z-y).
\end{align*}
Thus $(f_n)_{n\geq2}$ is uniformly equicontinuous on $[0,t]$.

The Arzel\`a--Ascoli theorem \cite{Rud76} now gives a uniformly convergent subsequence with continuous limit $f$.

Step~4. It remains to show that $f$ is of BV. As $f$ is the uniform limit of a subsequence of $(f_n)_{n\geq 2}$, the same uniform Lipschitz bound $B$ passes to $f$. Therefore, $V(f;[0,t])\leq Bt<\infty$. 

For later use, regularity also gives
\begin{align*}
V(f_n;[0,t])
&\leq\sum_{i=1}^{\min(K_t,n-1)}V(f_n;I_{i,n})\\
&\leq K_t g(2M).
\end{align*}
Passing to the limit in the variation sum for each fixed finite partition of $[0,t]$ yields $V(f;[0,t])\leq K_tg(2M)$ as well. This proves the claim.
\end{IEEEproof}

\begin{IEEEproof}[Proof of \cref{thm:sts}]
Step~1. By Step~1 of the proof of Theorem~3, since $\mathcal{P}$ is non-vanishing and has diminishing tails, for any fixed $t \in (0,1)$ there exists a finite integer $K_t \ge 1$ such that $[0,t] \subset \bigcup_{i=1}^{K_t} I_{i,n}$ for all $n > K_t$, and 
\begin{align*}
\inf_{n \ge 2, \, i \le \min(K_t, n-1)} |I_{i,n}| =: a > 0. 
\end{align*}
Since $\mathcal{P}$ is index-wise convergent, the endpoints $e_{i,n}$ converge as $n \to \infty$ to $e_i := \sum_{k=1}^{i-1} a_k$ with $e_{i+1} - e_i = a_i \ge a > 0$ for $1 \le i \le K_t$, and $t \le e_{K_t+1}$.

Step~2. Index the negative eigenvalues of $T_{\mathbf{W}}$ as $\lambda_{-1} < \lambda_{-2} < \dots < 0$ with orthonormal eigenfunctions $\psi_{-i} \in L^2([0,1])$, and let $\widehat{\mathbf{x}}(i) := \langle \psi_{-i}, x \rangle$. We claim that under \cref{ass:spec}, for every fixed $i \ge 1$,
\begin{equation}
\label{eq:gft-coord-conv}
\lim_{n \to \infty} \frac{|\widehat{\mathbf{x}_n}(i)|}{\sqrt{n}} = |\widehat{\mathbf{x}}(i)| \qquad \text{almost surely.}
\end{equation}
To prove (\ref{eq:gft-coord-conv}) for general continuous $x \in C([0,1])$, observe first that for any permutation $\pi$ of $\{1, \dots, n\}$ with permutation matrix $\mathbf{P}_\pi$, replacing $(\mathbf{A}_n, \mathbf{x}_n, \mathbf{u}_{i,n})$ by $(\mathbf{P}_\pi \mathbf{A}_n \mathbf{P}_\pi^\top, \mathbf{P}_\pi \mathbf{x}_n, \mathbf{P}_\pi \mathbf{u}_{i,n})$ preserves the eigenvalue ordering and leaves the inner product $\widehat{\mathbf{x}_n}(i) = \langle \mathbf{P}_\pi \mathbf{u}_{i,n}, \mathbf{P}_\pi \mathbf{x}_n \rangle = \langle \mathbf{u}_{i,n}, \mathbf{x}_n \rangle$ completely unchanged. We may therefore reorder the vertices of $G_n$ according to the order statistics of the latent sample points: $0 \le s_{(1)} < s_{(2)} < \dots < s_{(n)} \le 1$.

Partition $[0,1]$ into $n$ equal subintervals $J_{k,n} = [\frac{k-1}{n}, \frac{k}{n})$ ($1 \le k \le n$)\, including the right endpoint $1$ in $J_{n,n}$. In the ordered vertex labeling, let $\mathbf{W}_n : [0,1]^2 \to \{0,1\}$ be the step graphon induced by $\mathbf{A}_n$, and 
\begin{align*}
&x_n^\sharp(u) := \sum_{k=1}^n x(s_{(k)})\mathbf{1}_{J_{k,n}}(u); \\ &\psi_{-i,n}(u) := \sqrt{n}\sum_{k=1}^n \mathbf{u}_{i,n}(k)\mathbf{1}_{J_{k,n}}(u)
\end{align*}
be the step-function embedding of the sampled vector $\mathbf{x}_n$ and the normalized step eigenfunction of $T_{\mathbf{W}_n}$ corresponding to $\lambda_{i,n}/n$. According to \cite[Lemma 2]{Lua21}, the following holds
\begin{align*}
\frac{\widehat{\mathbf{x}_n}(i)}{\sqrt{n}} = \langle \psi_{-i,n}, x_n^\sharp \rangle.
\end{align*}

By the Glivenko-Cantelli theorem \cite{van98}, we have
\begin{align*}
\max_{1 \le k \le n} \big|s_{(k)} - \tfrac{k}{n}\big| \to 0 \text{ a.s.}
\end{align*}
Since $x$ is uniformly continuous on $[0,1]$, $\|x_n^\sharp - x\|_{\infty} \to 0$ and hence $\|x_n^\sharp - x\|_{L^2([0,1])} \to 0$ a.s. Furthermore, let $\overline{\mathbf{W}}_n(u,v) := \sum_{j,k=1}^n \mathbf{W}(s_{(j)}, s_{(k)})\mathbf{1}_{J_{j,n}}(u)\mathbf{1}_{J_{k,n}}(v)$. Approximating $\mathbf{W} \in L^1([0,1]^2)$ by continuous functions and applying the strong law for $U$-statistics to the off-diagonal empirical averages (the bounded diagonal contribution is $O(1/n)$), together with $\max_k |s_{(k)} - k/n| \to 0$ gives $\|\overline{\mathbf{W}}_n - \mathbf{W}\|_{L^1([0,1]^2)} \to 0$ a.s., while $\|\mathbf{W}_n - \overline{\mathbf{W}}_n\|_\square \to 0$ a.s.\ follows from a direct concentration argument. 

Conditionally on the latent positions, the edge indicators $\bA_{jk}$ for $j<k$ are independent Bernoulli variables with parameters $\mathbf{W}(s_{(j)},s_{(k)})$. Since the cut norm of a step kernel on the $n \times n$ block grid is attained on unions of blocks, we have (here: $1/n$ bounds the diagonal blocks)
\begin{align*}
&\big\|\mathbf{W}_n - \overline{\mathbf{W}}_n\big\|_\square \\ \le &\frac{1}{n} + \frac{1}{n^2}\max_{S,T \subseteq \{1,\dots,n\}} \Big|\sum_{j \in S,\, k \in T,\, j \ne k} (\bA_{jk} \\ & - \mathbf{W}(s_{(j)}, s_{(k)}))\Big|.
\end{align*}
For fixed $(S,T)$ the inner sum is a sum over unordered pairs of independent centred terms with ranges $c_{jk} \in \{0,1,2\}$ and $\sum c_{jk}^2 \le 2n^2$, so Hoeffding's inequality bounds the conditional probability that it exceeds $\epsilon n^2$ in absolute value by $2e^{-\epsilon^2 n^2}$. A union bound over the $4^n$ pairs $(S,T)$ and the Borel-Cantelli lemma then give $\limsup_{n} \|\mathbf{W}_n - \overline{\mathbf{W}}_n\|_\square \le \epsilon$ a.s.\ for every $\epsilon > 0$. Hence, $\|\mathbf{W}_n - \mathbf{W}\|_\square \to 0$ and 
\begin{align*}
\|T_{\mathbf{W}_n} - T_{\mathbf{W}}\|_{L^2 \to L^2} \le 2\sqrt{2}\|\mathbf{W}_n - \mathbf{W}\|_\square^{1/2} \to 0 \text{ a.s.} 
\end{align*}
Since $\lambda_{-i} < 0$ is a simple isolated eigenvalue of $T_{\mathbf{W}}$ (\cref{ass:spec}), by \cite[Proposition 2]{Lua21} there exist signs $\varepsilon_{i,n} \in \{-1, +1\}$ such that $\|\varepsilon_{i,n}\psi_{-i,n} - \psi_{-i}\|_{L^2([0,1])} \to 0$ a.s. The triangle and Cauchy-Schwarz inequalities imply
\begin{align*}
&\Big| \frac{|\widehat{\mathbf{x}_n}(i)|}{\sqrt{n}} - |\widehat{\mathbf{x}}(i)| \Big| \\
\le &\big| \langle \varepsilon_{i,n}\psi_{-i,n}, x_n^\sharp \rangle - \langle \psi_{-i}, x \rangle \big| \\
\le &\|\varepsilon_{i,n}\psi_{-i,n} - \psi_{-i}\|_{L^2([0,1])} \|x_n^\sharp\|_{L^2([0,1])} \\
&+ \|\psi_{-i}\|_{L^2([0,1])} \|x_n^\sharp - x\|_{L^2([0,1])} \xrightarrow{n \to \infty} 0 \text{ a.s.},
\end{align*}
proving \cref{eq:gft-coord-conv}. By the strong law of large numbers, we also have $\|\mathbf{x}_n\|_2/\sqrt{n} \to \|x\|_{L^2([0,1])}$ almost surely. Since a countable intersection of almost-sure events has probability $1$, all of these limits hold simultaneously on a single event $\Omega_0$ of probability $1$. Fix any realization in $\Omega_0$.

Step~3. Let $\widehat{\mathbf{x}}_t := (\widehat{\mathbf{x}}(1), \dots, \widehat{\mathbf{x}}(K_t+1))^\top$, $\mathcal{P}_\infty := \{I_1, \dots, I_{K_t}\}$ with $I_i = [e_i, e_{i+1}]$, and define $f := p(\mathcal{P}_\infty; |\widehat{\mathbf{x}}_t|)$ on $[0, e_{K_t+1}] \supset [0,t]$.
\begin{itemize}[leftmargin=1.5em]
    \item If $y \in [0,t] \setminus \{e_1, \dots, e_{K_t+1}\}$, then $y \in (e_i, e_{i+1})$ for a unique $i \le K_t$. Since $e_{i,n} \to e_i$ and $e_{i+1,n} \to e_{i+1}$, there exists $N_y$ such that $y \in (e_{i,n}, e_{i+1,n})$ for all $n \ge N_y$. By consistency of the interpolation and continuity of $\phi$,
    \begin{align*}
    \lim_{n \to \infty} f_n(y) &= \lim_{n \to \infty} \phi\!\left(\frac{|\widehat{\mathbf{x}_n}(i)|}{\sqrt{n}}, \frac{|\widehat{\mathbf{x}_n}(i+1)|}{\sqrt{n}}, \frac{y - e_{i,n}}{|I_{i,n}|}\right) \\
    &= \phi\!\left(|\widehat{\mathbf{x}}_t(i)|, |\widehat{\mathbf{x}}_t(i+1)|, \frac{y - e_i}{|I_i|}\right) = f(y).
    \end{align*}
    \item If $y = e_i$ for some $1 \le i \le K_t + 1$, then for all sufficiently large $n$, $e_i$ belongs to either $I_{i-1,n}$ (when $e_i \le e_{i,n}$, with $i \ge 2$) or $I_{i,n}$ (when $e_i \ge e_{i,n}$, with $i \le K_t$; for $i = K_t + 1$, which matters only if $e_{K_t+1} = t$, Step~1 gives $e_{K_t+1,n} \ge t = e_i$ for $n > K_t$, so the first case applies). Along any subsequence $n_j$ where $e_i \in I_{i-1, n_j}$, we have $\frac{e_i - e_{i-1, n_j}}{|I_{i-1, n_j}|} \to 1$, so
    \begin{align*}
    \lim_{j \to \infty} f_{n_j}(e_i) & = \phi(|\widehat{\mathbf{x}}_t(i-1)|, |\widehat{\mathbf{x}}_t(i)|, 1) \\ & = |\widehat{\mathbf{x}}_t(i)| = f(e_i).
    \end{align*}
    Along any subsequence $n_k$ where $e_i \in I_{i, n_k}$, we have $\frac{e_i - e_{i, n_k}}{|I_{i, n_k}|} \to 0$, so 
    \begin{align*}
    \lim_{k \to \infty} f_{n_k}(e_i) & = \phi(|\widehat{\mathbf{x}}_t(i)|, |\widehat{\mathbf{x}}_t(i+1)|, 0) \\ &= |\widehat{\mathbf{x}}_t(i)| = f(e_i). 
    \end{align*}
    Thus $\lim_{n \to \infty} f_n(e_i) = f(e_i)$.
\end{itemize}

Step~4. On $\Omega_0$, $M:=\sup_{n\geq2}\|\mathbf{x}_n\|_2/\sqrt{n}<\infty$. By Step~1 and regularity, each relevant interpolation interval has Lipschitz constant at most $g(2M)/a$. Adding the bounds across interval endpoints gives the same Lipschitz constant on $[0,t]$, uniformly in $n$. Moreover,
\[
\sup_{n\geq2}V(f_n;[0,t])\leq K_tg(2M).
\]
Uniform equicontinuity and the pointwise convergence from Step~3 imply uniform convergence on $[0,t]$. Passing to the limit in finite variation sums gives $V(f;[0,t])\leq K_tg(2M)$.

Finally, to justify the $L^2$ extension, choose continuous $g_m$ with $\|x-g_m\|_{L^2}\to0$, and let $g_{m,n}^\sharp$ be their sorted step embeddings. For each fixed $m$, the strong law gives
\[
\|x_n^\sharp-g_{m,n}^\sharp\|_{L^2}^2
=\frac1n\sum_{i=1}^n|x(s_i)-g_m(s_i)|^2
\longrightarrow\|x-g_m\|_{L^2}^2
\]
almost surely. The continuous case gives $g_{m,n}^\sharp\to g_m$ in $L^2$. On a single probability-one event for all $m$, the triangle inequality yields $\limsup_n\|x_n^\sharp-x\|_{L^2}\leq2\|x-g_m\|_{L^2}$, hence $x_n^\sharp\to x$ in $L^2$. The norm strong law and the remainder of the proof apply unchanged.
\end{IEEEproof}

\bibliographystyle{IEEEtran}
\bibliography{IEEEabrv,StringDefinitions,ref}

\end{document}

%% file: preamble.tex
\usepackage[T1]{fontenc}
\usepackage[utf8]{inputenc}
\usepackage{mathtools}

\usepackage{amssymb,mathrsfs}
\usepackage{amsthm}
\usepackage{bm}
\usepackage{scalerel}
\usepackage{nicefrac}
\usepackage{microtype} 
\usepackage[shortlabels]{enumitem}
\usepackage{graphicx}
\usepackage{epstopdf}
\DeclareGraphicsExtensions{.eps,.png,.jpg,.pdf}

\usepackage{url}
\usepackage{colortbl}
\usepackage{booktabs}
\usepackage{multirow}
\usepackage{colortbl,xcolor}
\usepackage{soul}
\usepackage{xparse,xstring}
\usepackage{calc}
\usepackage{etoolbox}

\makeatletter
\@ifpackageloaded{natbib}{
	\relax
}{
	\usepackage{cite}
}
\makeatother

\usepackage{array}
\newcolumntype{L}[1]{>{\raggedright\let\newline\\\arraybackslash\hspace{0pt}}m{#1}}
\newcolumntype{C}[1]{>{\centering\let\newline\\\arraybackslash\hspace{0pt}}m{#1}}
\newcolumntype{R}[1]{>{\raggedleft\let\newline\\\arraybackslash\hspace{0pt}}m{#1}}

\makeatletter
\let\MYcaption\@makecaption
\makeatother
\usepackage[font=footnotesize]{subcaption}
\makeatletter
\let\@makecaption\MYcaption
\makeatother

\usepackage{glossaries}
\makeatletter
\sfcode`\.1006

\let\oldgls\gls
\let\oldglspl\glspl

\newcommand\fussy@ifnextchar[3]{%
	\let\reserved@d=#1%
	\def\reserved@a{#2}%
	\def\reserved@b{#3}%
	\futurelet\@let@token\fussy@ifnch}
\def\fussy@ifnch{%
	\ifx\@let@token\reserved@d
		\let\reserved@c\reserved@a
	\else
		\let\reserved@c\reserved@b
	\fi
	\reserved@c}

\renewcommand{\gls}[1]{%
\oldgls{#1}\fussy@ifnextchar.{\@checkperiod}{\@}}
\renewcommand{\glspl}[1]{%
\oldglspl{#1}\fussy@ifnextchar.{\@checkperiod}{\@}}

\newcommand{\@checkperiod}[1]{%
	\ifnum\sfcode`\.=\spacefactor\else#1\fi
}

\robustify{\gls}
\robustify{\glspl}
\makeatother

\newacronym{wrt}{w.r.t.}{with respect to}
\newacronym{RHS}{R.H.S.}{right-hand side}
\newacronym{LHS}{L.H.S.}{left-hand side}
\newacronym{iid}{i.i.d.}{independent and identically distributed}
\newacronym{SOTA}{SOTA}{state-of-the-art}

\usepackage{float}

\ifx\notloadhyperref\undefined
	\ifx\loadbibentry\undefined
		\usepackage[hidelinks,hypertexnames=false]{hyperref}
	\else
		\usepackage{bibentry}
		\makeatletter\let\saved@bibitem\@bibitem\makeatother
		\usepackage[hidelinks,hypertexnames=false]{hyperref}
		\makeatletter\let\@bibitem\saved@bibitem\makeatother
	\fi
\else
	\ifx\loadbibentry\undefined
		\relax
	\else
		\usepackage{bibentry}
	\fi
\fi

\usepackage{cleveref-forward}

\crefname{equation}{}{}
\Crefname{equation}{}{}
\crefname{claim}{claim}{claims}
\crefname{step}{step}{steps}
\crefname{line}{line}{lines}
\crefname{condition}{condition}{conditions}
\crefname{dmath}{}{}
\crefname{dseries}{}{}
\crefname{dgroup}{}{}
\crefname{page}{page}{pages}

\crefname{Problem}{Problem}{Problems}
\crefformat{Problem}{Problem~#2#1#3}
\crefrangeformat{Problem}{Problems~#3#1#4 to~#5#2#6}

\crefname{Theorem}{Theorem}{Theorems}
\crefname{Corollary}{Corollary}{Corollaries}
\crefname{Proposition}{Proposition}{Propositions}
\crefname{Lemma}{Lemma}{Lemmas}
\crefname{Definition}{Definition}{Definitions}
\crefname{Example}{Example}{Examples}
\crefname{Assumption}{Assumption}{Assumptions}
\crefname{Remark}{Remark}{Remarks}
\crefname{Rem}{Remark}{Remarks}
\crefname{remarks}{Remarks}{Remarks}
\crefname{Appendix}{Appendix}{Appendices}
\crefname{Supplement}{Supplement}{Supplements}
\crefname{Exercise}{Exercise}{Exercises}
\crefname{TheoremA}{Theorem}{Theorems}
\crefname{CorollaryA}{Corollary}{Corollaries}
\crefname{PropositionA}{Proposition}{Propositions}
\crefname{LemmaA}{Lemma}{Lemmas}
\crefname{DefinitionA}{Definition}{Definitions}
\crefname{ExampleA}{Example}{Examples}
\crefname{RemarkA}{Remark}{Remarks}
\crefname{AssumptionA}{Assumption}{Assumptions}
\crefname{ExerciseA}{Exercise}{Exercises}
\crefname{algorithm}{Algorithm}{Algorithms}
\crefname{figure}{Fig.}{Figs.}
\crefname{table}{Table}{Tables}
\crefname{section}{Section}{Sections}
\crefname{subsection}{Section}{Sections}
\crefname{subsubsection}{Section}{Sections}

\usepackage{crossreftools}
\ifx\notloadhyperref\undefined
\else
	\relax
\fi

\usepackage{algorithm}
\usepackage{algpseudocode}

\ifx\loadbreqn\undefined
	\relax
\else
	\usepackage{breqn}
\fi

\makeatletter
\def\cleartheorem#1{%
    \expandafter\let\csname#1\endcsname\relax
    \expandafter\let\csname c@#1\endcsname\relax
}
\def\clearthms#1{ \@for\tname:=#1\do{\cleartheorem\tname} }
\makeatother

\ifx\renewtheorem\undefined
	\ifx\useTheoremCounter\undefined
		\newtheorem{Theorem}{Theorem}
		\newtheorem{Corollary}{Corollary}
		\newtheorem{Proposition}{Proposition}
		
	\else
		\newtheorem{Theorem}{Theorem}

	\fi

	\newtheorem{Definition}{Definition}
	\newtheorem{Example}{Example}
	\newtheorem{Remark}{Remark}
	\newtheorem{Assumption}{Assumption}
	
	\newtheorem{Problem}{Problem}

\fi

\theoremstyle{remark}

\theoremstyle{plain}

\newcommand{\qednew}{\nobreak \ifvmode \relax \else
		\ifdim\lastskip<1.5em \hskip-\lastskip
			\hskip1.5em plus0em minus0.5em \fi \nobreak
		\vrule height0.75em width0.5em depth0.25em\fi}



\NewDocumentCommand{\movedownsub}{e{^_}}{%
	\IfNoValueTF{#1}{%
		\IfNoValueF{#2}{^{}}
	}{%
		^{#1}
	}%
	\IfNoValueF{#2}{_{#2}}
}

\let\latexchi\chi
\RenewDocumentCommand{\chi}{}{\latexchi\movedownsub}

\newcommand{\ba}{\mathbf{a}}
\newcommand{\bA}{\mathbf{A}}

\newcommand{\boldf}{\mathbf{f}}

\newcommand{\bI}{\mathbf{I}}

\newcommand{\bL}{\mathbf{L}}

\newcommand{\bs}{\mathbf{s}}
\newcommand{\bS}{\mathbf{S}}

\newcommand{\bu}{\mathbf{u}}
\newcommand{\bU}{\mathbf{U}}

\newcommand{\bW}{\mathbf{W}}
\newcommand{\bx}{\mathbf{x}}
\newcommand{\bX}{\mathbf{X}}
\newcommand{\by}{\mathbf{y}}

\newcommand{\bz}{\mathbf{z}}

\DeclareSymbolFont{bsfletters}{OT1}{cmss}{bx}{n}
\DeclareSymbolFont{ssfletters}{OT1}{cmss}{m}{n}
\DeclareMathSymbol{\bsfGamma}{0}{bsfletters}{'000}
\DeclareMathSymbol{\ssfGamma}{0}{ssfletters}{'000}
\DeclareMathSymbol{\bsfDelta}{0}{bsfletters}{'001}
\DeclareMathSymbol{\ssfDelta}{0}{ssfletters}{'001}
\DeclareMathSymbol{\bsfTheta}{0}{bsfletters}{'002}
\DeclareMathSymbol{\ssfTheta}{0}{ssfletters}{'002}
\DeclareMathSymbol{\bsfLambda}{0}{bsfletters}{'003}
\DeclareMathSymbol{\ssfLambda}{0}{ssfletters}{'003}
\DeclareMathSymbol{\bsfXi}{0}{bsfletters}{'004}
\DeclareMathSymbol{\ssfXi}{0}{ssfletters}{'004}
\DeclareMathSymbol{\bsfPi}{0}{bsfletters}{'005}
\DeclareMathSymbol{\ssfPi}{0}{ssfletters}{'005}
\DeclareMathSymbol{\bsfSigma}{0}{bsfletters}{'006}
\DeclareMathSymbol{\ssfSigma}{0}{ssfletters}{'006}
\DeclareMathSymbol{\bsfUpsilon}{0}{bsfletters}{'007}
\DeclareMathSymbol{\ssfUpsilon}{0}{ssfletters}{'007}
\DeclareMathSymbol{\bsfPhi}{0}{bsfletters}{'010}
\DeclareMathSymbol{\ssfPhi}{0}{ssfletters}{'010}
\DeclareMathSymbol{\bsfPsi}{0}{bsfletters}{'011}
\DeclareMathSymbol{\ssfPsi}{0}{ssfletters}{'011}
\DeclareMathSymbol{\bsfOmega}{0}{bsfletters}{'012}
\DeclareMathSymbol{\ssfOmega}{0}{ssfletters}{'012}

\makeatletter
\newcommand*\rel@kern[1]{\kern#1\dimexpr\macc@kerna}
\newcommand*\widebar[1]{%
  \begingroup
  \def\mathaccent##1##2{%
    \rel@kern{0.8}%
    \overline{\rel@kern{-0.8}\macc@nucleus\rel@kern{0.2}}%
    \rel@kern{-0.2}%
  }%
  \macc@depth\@ne
  \let\math@bgroup\@empty \let\math@egroup\macc@set@skewchar
  \mathsurround\z@ \frozen@everymath{\mathgroup\macc@group\relax}%
  \macc@set@skewchar\relax
  \let\mathaccentV\macc@nested@a
  \macc@nested@a\relax111{#1}%
  \endgroup
}
\makeatother

\DeclareMathOperator{\var}{var}

\DeclareMathOperator{\cov}{cov}

\newcommand{\ifbcdot}[1]{\ifblank{#1}{\cdot}{#1}}

\DeclarePairedDelimiterX\abs[1]{\lvert}{\rvert}{\ifbcdot{#1}}
\DeclarePairedDelimiterX\parens[1]{(}{)}{\ifbcdot{#1}}
\DeclarePairedDelimiterX\brk[1]{[}{]}{\ifbcdot{#1}}
\DeclarePairedDelimiterX\braces[1]{\{}{\}}{\ifbcdot{#1}}
\DeclarePairedDelimiterX\angles[1]{\langle}{\rangle}{\ifblank{#1}{\cdot,\cdot}{#1}}
\DeclarePairedDelimiterX\ip[2]{\langle}{\rangle}{\ifbcdot{#1},\ifbcdot{#2}}
\DeclarePairedDelimiterX\norm[1]{\lVert}{\rVert}{\ifbcdot{#1}}
\DeclarePairedDelimiterX\ceil[1]{\lceil}{\rceil}{\ifbcdot{#1}}
\DeclarePairedDelimiterX\floor[1]{\lfloor}{\rfloor}{\ifbcdot{#1}}

\DeclareFontFamily{U}{matha}{\hyphenchar\font45}
\DeclareFontShape{U}{matha}{m}{n}{
      <5> <6> <7> <8> <9> <10> gen * matha
      <10.95> matha10 <12> <14.4> <17.28> <20.74> <24.88> matha12
      }{}
\DeclareSymbolFont{matha}{U}{matha}{m}{n}
\DeclareFontSubstitution{U}{matha}{m}{n}

\DeclareFontFamily{U}{mathx}{\hyphenchar\font45}
\DeclareFontShape{U}{mathx}{m}{n}{
      <5> <6> <7> <8> <9> <10>
      <10.95> <12> <14.4> <17.28> <20.74> <24.88>
      mathx10
      }{}
\DeclareSymbolFont{mathx}{U}{mathx}{m}{n}
\DeclareFontSubstitution{U}{mathx}{m}{n}

\DeclareMathDelimiter{\vvvert}{0}{matha}{"7E}{mathx}{"17}
\DeclarePairedDelimiterX\vertiii[1]{\vvvert}{\vvvert}{\ifbcdot{#1}}

\DeclarePairedDelimiterXPP\trace[1]{\operatorname{Tr}}{(}{)}{}{\ifbcdot{#1}} 
\DeclarePairedDelimiterXPP\col[1]{\operatorname{col}}{\{}{\}}{}{\ifbcdot{#1}} 
\DeclarePairedDelimiterXPP\row[1]{\operatorname{row}}{\{}{\}}{}{\ifbcdot{#1}} 
\DeclarePairedDelimiterXPP\erf[1]{\operatorname{erf}}{(}{)}{}{\ifbcdot{#1}}
\DeclarePairedDelimiterXPP\erfc[1]{\operatorname{erfc}}{(}{)}{}{\ifbcdot{#1}}
\DeclarePairedDelimiterXPP\KLD[2]{D}{(}{)}{}{\ifbcdot{#1}\, \delimsize\|\, \ifbcdot{#2}} 
\DeclarePairedDelimiterXPP\op[2]{\operatorname{#1}}{(}{)}{}{#2} 

\newcommand{\ud}{\,\mathrm{d}} 

\DeclarePairedDelimiterXPP\indicate[1]{{\bf 1}}{\{}{\}}{}{\ifbcdot{#1}}

\NewDocumentCommand\ofrac{s m}{%
	\IfBooleanTF#1%
	{\dfrac{1}{#2}}%
	{\frac{1}{#2}}%
}
\NewDocumentCommand\ddfrac{s m m}{%
	\IfBooleanTF#1%
	{\dfrac{\mathrm{d} {#2}}{\mathrm{d} {#3}}}%
	{\frac{\mathrm{d} {#2}}{\mathrm{d} {#3}}}%
}
\NewDocumentCommand\ppfrac{s m m}{%
	\IfBooleanTF#1%
	{\dfrac{\partial {#2}}{\partial {#3}}}%
	{\frac{\partial {#2}}{\partial {#3}}}%
}

\newcommand{\setgiven}{:}
\providecommand\given{}

\DeclarePairedDelimiterX\Set[2]\{\}{%
	\if#1:%
		\renewcommand\given{\SetSymbol{:}}%
	\else%
		\renewcommand\given{\SetSymbol[\delimsize]{#1}}%
	\fi%
#2
}

\NewDocumentCommand\set{s O{\setgiven} m}{%
	\IfBooleanTF#1%
	{\Set*{#2}{#3}}%
	{\Set{#2}{#3}}%
}

\NewDocumentCommand{\evalat}{ s O{\big} m e{_^} }{%
\IfBooleanTF{#1}%
{\left. #3 \right|}{#3#2|}%
\IfValueT{#4}{_{#4}}%
\IfValueT{#5}{^{#5}}%
}

\providecommand\given{}
\DeclarePairedDelimiterXPP\cprob[1]{}(){}{
\renewcommand\given{\nonscript\,\delimsize\vert\allowbreak\nonscript\,\mathopen{}}%
#1%
}
\DeclarePairedDelimiterXPP\cexp[1]{}[]{}{
\renewcommand\given{\nonscript\,\delimsize\vert\allowbreak\nonscript\,\mathopen{}}%
#1%
}

\DeclareDocumentCommand \P { s e{_^} d() g } {%
	\mathbb{P}%
	\IfBooleanTF{#1}%
		{
			\IfValueT{#2}{_{#2}}%
			\IfValueT{#3}{^{#3}}%
			\IfValueTF{#5}{\cprob{#4 \given #5}}{\IfValueT{#4}{\cprob{#4}}}%
		}%
		{
			\IfValueT{#2}{_{#2}}%
			\IfValueT{#3}{^{#3}}%
			\IfValueTF{#5}{\cprob*{#4 \given #5}}{\IfValueT{#4}{\cprob*{#4}}}%
		}%
}

\DeclareDocumentCommand \E { s e{_^} o g } {%
	\mathbb{E}%
	\IfBooleanTF{#1}%
		{
			\IfValueT{#2}{_{#2}}%
			\IfValueT{#3}{^{#3}}%
			\IfValueTF{#5}{\cexp{#4 \given #5}}{\IfValueT{#4}{\cexp{#4}}}%
		}%
		{
			\IfValueT{#2}{_{#2}}%
			\IfValueT{#3}{^{#3}}%
			\IfValueTF{#5}{\cexp*{#4 \given #5}}{\IfValueT{#4}{\cexp*{#4}}}%
		}%
}

\DeclareDocumentCommand \Var { s e{_^} d() g } {%
	\var%
	\IfBooleanTF{#1}%
		{
			\IfValueT{#2}{_{#2}}%
			\IfValueT{#3}{^{#3}}%
			\IfValueTF{#5}{\cprob{#4 \given #5}}{\IfValueT{#4}{\cprob{#4}}}%
		}%
		{
			\IfValueT{#2}{_{#2}}%
			\IfValueT{#3}{^{#3}}%
			\IfValueTF{#5}{\cprob*{#4 \given #5}}{\IfValueT{#4}{\cprob*{#4}}}%
		}%
}

\DeclareDocumentCommand \Cov { s e{_^} d() g } {%
	\cov%
	\IfBooleanTF{#1}%
		{
			\IfValueT{#2}{_{#2}}%
			\IfValueT{#3}{^{#3}}%
			\IfValueTF{#5}{\cprob{#4 \given #5}}{\IfValueT{#4}{\cprob{#4}}}%
		}%
		{
			\IfValueT{#2}{_{#2}}%
			\IfValueT{#3}{^{#3}}%
			\IfValueTF{#5}{\cprob*{#4 \given #5}}{\IfValueT{#4}{\cprob*{#4}}}%
		}%
}

\ExplSyntaxOn
\NewDocumentCommand \dist {m o o} {%
\mathrm{#1}\left(%
	\IfValueT{#3}{%
		\tl_if_blank:nTF{ #3 }{\cdot\, \middle|\, }{#3\, \middle|\, }%
	}
	\IfValueT{#2}{#2}%
\right)%
}
\ExplSyntaxOff

\NewDocumentCommand {\cbrace} {t+ D[]{black} D(){\widthof{#5}} m m } {%
	\begingroup%
		\color{#2}
		\IfBooleanTF{#1}{%
			\overbrace{#4}^%
		}{
			\underbrace{#4}_%
		}%
		{\parbox[c]{#3}{\centering\footnotesize{#5}}}%
	\endgroup%
}

\let\oldforall\forall
\renewcommand{\forall}{\oldforall \, }

\let\oldexist\exists
\renewcommand{\exists}{\oldexist \, }

\makeatletter

\newcommand{\rankcolor}[2]{%
	\expandafter\renewcommand\csname #1\endcsname[1]{%
		\ifblank{##1}{%
			{\color{#2} \textbf{#2}}%
		}{%
			\ifmmode
				\textcolor{#2}{\bm{##1}}%
			\else%
				{\color{#2} \textbf{##1}}%
			\fi	
		}%
	}
}

\rankcolor{first}{red}
\rankcolor{second}{blue}
\rankcolor{third}{cyan}
\makeatother

\graphicspath{{./Figures/}{./figures/}}
\DeclareDocumentCommand{\includeCroppedPdf}{ o O{./Figures/} m }{
	\IfFileExists{#2#3-crop.pdf}{}{%
		\immediate\write18{pdfcrop #2#3.pdf #2#3-crop.pdf}}%
	\includegraphics[#1]{#2#3-crop.pdf}
}

\makeatletter
\newcommand*{\addFileDependency}[1]{
  \typeout{(#1)}
  \@addtofilelist{#1}
  \IfFileExists{#1}{}{\typeout{No file #1.}}
}
\makeatother

\definecolor{gray90}{gray}{0.9}
\def\colorlist{red,blue,brown,cyan,darkgray,gray,lightgray,green,lime,magenta,olive,orange,pink,purple,teal,violet,white,yellow}

\makeatletter
\def\startcomment{[}
\ifx\nohighlights\undefined
	\newcommand{\createcolor}[1]{%
			\expandafter\newcommand\csname #1\endcsname[1]{{\color{#1} ##1}}%
	}
	\newcommand{\msout}[1]{\text{\color{green} \st{\ensuremath{#1}}}}
	\newcommand{\del}[1]{{\color{green}\ifmmode \msout{#1}\else\st{#1}\fi}}
\else
	\newcommand{\createcolor}[1]{%
			\expandafter\newcommand\csname #1\endcsname[1]{%
				\noexpandarg%
				\StrChar{##1}{1}[\firstletter]%
				\if\firstletter\startcomment%
					\relax
				\else%
					##1
				\fi
			}%
	}
	\newcommand{\msout}[1]{}
	\newcommand{\del}[1]{}
\fi

\def\@tempa#1,{%
    \ifx\relax#1\relax\else
        \createcolor{#1}%
        \expandafter\@tempa
    \fi
}
\expandafter\@tempa\colorlist,\relax,
\makeatother

\newcommand{\hhide}[1]{}

\ifx\diagnoselabel\undefined
	\relax
\else
	\makeatletter
	\def\@testdef #1#2#3{%
		\def\reserved@a{#3}\expandafter \ifx \csname #1@#2\endcsname
			\reserved@a  \else
			\typeout{^^Jlabel #2 changed:^^J%
				\meaning\reserved@a^^J%
				\expandafter\meaning\csname #1@#2\endcsname^^J}%
			\@tempswatrue \fi}
	\makeatother
\fi


%% file: math_commands.tex
\usepackage{amsmath,amsfonts,bm}

\def\eqref#1{equation~\ref{#1}}

\def\1{\bm{1}}

\DeclareMathAlphabet{\mathsfit}{\encodingdefault}{\sfdefault}{m}{sl}
\SetMathAlphabet{\mathsfit}{bold}{\encodingdefault}{\sfdefault}{bx}{n}

\newcommand{\R}{\mathbb{R}}



%% file: IEEEabrv.bib
@STRING{IEEE_J_SP         = "{IEEE} Trans. Signal Process."}

@STRING{IEEE_J_PROC       = "Proc. {IEEE}"}

@STRING{IEEE_M_SP         = "{IEEE} Signal Process. Mag."}


%% file: StringDefinitions.bib
@STRING{ICASSP = {Proc. IEEE Int. Conf. Acoustics, Speech, and Signal Processing}}

@STRING{IEEE = {The Institute of Electrical and Electronics Engineers}}


%% file: ref.bib
@article{Cec20,
  author  = {E. Ceci and S. Barbarossa},
  title   = {Graph signal processing in the presence of topology uncertainties},
  journal = {IEEE Trans. Signal Process.},
  year    = {2020},
  volume  = {68},
  number  = {},
  pages   = {1558–1573},
}

@article{Jan:J13,
  title   = {Graphons, cut norm and distance, couplings and rearrangements},
  journal = {N. Y. J. Math.},
  volume  = {4},
  number  = {},
  pages   = {},
  year    = {2013},
  author  = {Svante Janson},
}

@article{Ji19,
  author  = {F. Ji and W. P. Tay},
  title   = {A {Hilbert} Space Theory of Generalized Graph Signal Processing},
  journal = {{IEEE} Trans. Signal Process.},
  year    = {2019},
  volume  = {67},
  number  = {24},
  pages   = {6188 -- 6203},
}

@article{Ji22,
  author  = {F. Ji and W. P. Tay and A. Ortega},
  title   = {Graph Signal Processing over a Probability Space of Shift Operators},
  journal = {IEEE Trans. Signal Process.},
  year    = {2023},
  volume  = {71},
  number  = {},
  pages   = {1159-1174},
}

@article{JiaFenTay:J23,
  year    = {2023},
  volume  = {17},
  journal = {Foundations and Trends® in Signal Processing},
  title   = {Generalizing Graph Signal Processing: High Dimensional Spaces, Models and Structures},
  number  = {3},
  pages   = {209-290},
  author  = {X. Jian and F. Ji and W. P. Tay},
}

@article{LeuMarMou:J23,
  author  = {Leus, G. and Marques, A. and Moura, Jo. and Ortega, A. and Shuman, D.},
  journal = IEEE_M_SP,
  title   = {Graph Signal Processing: History, development, impact, and outlook},
  year    = {2023},
  month   = jun,
  volume  = {40},
  number  = {4},
  pages   = {49-60},
}

@book{Lov:12,
  author    = {Lov{\'a}sz, L{\'a}szl{\'o}},
  publisher = {American Mathematical Society},
  title     = {Large Networks and Graph Limits},
  year      = {2012},
  address   = {Providence, RI, USA},
}

@article{Lua21,
  author  = {Ruiz, L. and Chamon, L. and Ribeiro, A.},
  journal = {{IEEE} Trans. Signal Process.},
  title   = {Graphon Signal Processing},
  volume  = {69},
  number  = {},
  pages   = {4961-4976},
  year    = {2021},
}

@article{OrtFroKov:J18,
  author  = {Ortega, A. and Frossard, P. and Kovacevic, J. and Moura, J. and Vandergheynst, P.},
  journal = IEEE_J_PROC,
  title   = {Graph Signal Processing: Overview, Challenges, and Applications},
  year    = {2018},
  volume  = {106},
  number  = {5},
  pages   = {808-828},
}

@inproceedings{Rui20,
  author    = {L. Ruiz and L. Chamon and A. Ribeiro},
  title     = {The graphon {F}ourier transform},
  booktitle = {ICASSP},
  year      = {2020},
}

@article{San13,
  author  = {A. Sandryhaila and J. M. F. Moura},
  title   = {Discrete signal processing on graphs},
  journal = {IEEE Trans. Signal Process.},
  year    = {2013},
  volume  = {61},
  number  = {7},
  pages   = {1644-1656},
}

@article{Seg17,
  author  = {S. Segarra and A. Marques and A. Ribeiro},
  title   = {Optimal graph-filter design and applications to distributed linear network
             operators},
  journal = {IEEE Trans. Signal Process.},
  year    = {2017},
  volume  = {65},
  number  = {15},
  pages   = {4117–4131},
}

@article{Shu13,
  author  = {D. Shuman and S. Narang and P. Frossard and A. Ortega and P. Vandergheynst},
  title   = {The emerging field of signal processing on graphs: Extending high-dimensional data analysis to networks and other irregular domains},
  journal = {IEEE Signal Process. Mag.},
  year    = {2013},
  volume  = {30},
  number  = {3},
  pages   = {83-98},
}

@article{Tsi15,
  author  = {M. {Tsitsvero} and S. {Barbarossa} and P. {Di Lorenzo}},
  title   = {Signals on Graphs: Uncertainty Principle and Sampling},
  journal = IEEE_J_SP,
  year    = {2016},
  volume  = {64},
  number  = {18},
  pages   = {4845-4860},
}

@book{Rud76,
  author    = {W. Rudin},
  title     = {Principles of Mathematical Analysis},
  publisher = {McGraw-Hill, Inc.},
  edition = {3rd ed.},
  year      = 1976,
}

@inproceedings{Def16,
  author = {M. Defferrard and X. Bresson and P. Vandergheynst},
  title = {Convolutional neural networks on graphs with fast localized spectral filtering},
  booktitle = {Advances in Neural Information Processing Systems},
  pages = {3844–3852},
  year = {2016}
}

@inproceedings{Kip17,
  author = {T. Kipf and M. Welling},
  title = {Semi-supervised classification with graph convolutional networks},
  booktitle = {Proc. International Conference on Learning Representations},
  year = {2017}
}

@ARTICLE{JiZhao25,
  author={Ji, Feng and Zhao, Yanan and Lee, See Hian and Zhao, Kai and Tay, Wee Peng and Yang, Jielong},
  journal={IEEE Trans. Signal Inf. Process. Netw. }, 
  title={Graph Distributional Signals for Regularization in Graph Neural Networks}, 
  year={2025},
  volume={11},
  number={},
  pages={670-682},
  }

@ARTICLE{JiJ24,
  author={Ji, Feng and Jian, Xingchao and Tay, Wee Peng},
  journal={IEEE Transactions on Signal Processing}, 
  title={Modeling Sparse Graph Sequences and Signals Using Generalized Graphons}, 
  year={2024},
  volume={72},
  number={},
  pages={5048-5064},
  }

@book{Ort22,
  title={Introduction to graph signal processing},
  author={Ortega, Antonio},
  year={2022},
  publisher={Cambridge University Press}
}

@article{Puy18,
Author = {Gilles Puy and Nicolas Tremblay and Rémi Gribonval and Pierre Vandergheynst},
title = {Random sampling of bandlimited signals on graphs},
journal = {Appl. Comput. Harmon. Anal.},
Volume = {44}, 
Issue = {2},
year = {2018},
Pages ={446-475},
}

@article{Lyo14,
  author  = {Terry Lyons},
  journal = {arXiv preprint 	arXiv:1405.4537},
  title   = {Rough paths, Signatures and the modelling of functions on streams},
  year    = {2014},
}

@incollection{Che26,
  author    = {Chevyrev, Ilya and Kormilitzin, Andrey},
  title     = {A Primer on the Signature Method in Machine Learning},
  booktitle = {Signature Methods in Finance: An Introduction with Computational Applications},
  editor    = {Bayer, Christian and dos Reis, Gon{\c{c}}alo and Horvath, Blanka and Oberhauser, Harald},
  series    = {Springer Finance},
  publisher = {Springer},
  address   = {Cham},
  year      = {2026},
  chapter   = {1},
  pages     = {1--72}
}

@article{Lev16,
  author  = {Daniel Levin and Terry Lyons and Hao Ni},
  journal = {arXiv preprint 		arXiv:1309.0260},
  title   = {Learning from the past, predicting the statistics for the future, learning an evolving system
},
  year    = {2016},
}

@book{Fri12,
  title={Multidimensional Stochastic Processes as Rough Paths},
  author={Peter Friz and Nicolas Victoir},
  year={2012},
  publisher={Cambridge University Press}
}

@article{Lev40,
author = {P. Levy},
title = {Le Mouvement Brownien Plan},
journal = {Amer. J. Math.},
volume = {62},
number = {1},
pages = {487–550},
year = {1940}}

@book{Dud99,
  title={Uniform Central Limit Theorems.},
  author={R. Dudley},
  year={1999},
  publisher={Cambridge University Press}
}

@misc{Mou25,
  author       = {Moustafa Mohamed},
  title        = {A Comprehensive Guide to Supervised Learning Models in Machine Learning},
  howpublished = {GitHub Community Discussions},
  year         = {2025},
  url          = {https://github.com/orgs/community/discussions/163632},
}

@article{Fly19,
  title={Change Detection with the Kernel Cumulative Sum Algorithm},
  author={Thomas Flynn and Shinjae Yoo},
  journal={IEEE Conference on Decision and Control},
  year={2019},
  pages={6092-6099},
}

@article{Arl19,
  TITLE = {A Kernel Multiple Change-point Algorithm via Model Selection},
  AUTHOR = {Arlot, Sylvain and Celisse, Alain and Harchaoui, Zaid},
  JOURNAL = {J. Mach. Learn. Res.},
  VOLUME = {20},
  NUMBER = {162},
  PAGES = {1--56},
  YEAR = {2019},
}

@article{Ngu21,
author = {Nguyen, Dai Hai and Nguyen, Canh Hao and Mamitsuka, Hiroshi},
title = {Learning subtree pattern importance for Weisfeiler-Lehman based graph kernels},
year = {2021},
volume = {110},
number = {7},
journal = {Mach. Learn.},
month = jul,
pages = {1585–1607},
numpages = {23},
}

@inproceedings{Mor20,
    title={{TUDataset}: A collection of benchmark datasets for learning with graphs},
    author={Christopher Morris and Nils M. Kriege and Franka Bause and Kristian Kersting and Petra Mutzel and Marion Neumann},
    booktitle={ICML Workshop on Graph Representation Learning and Beyond },
    pages={},
    year={2020}
}

@article{Fra19,
  author  = {Franz J. Kiraly and Harald Oberhauser},
  title   = {Kernels for Sequentially Ordered Data},
  journal = {J. Mach. Learn. Res.},
  year    = {2019},
  volume  = {20},
  number  = {31},
  pages   = {1--45},
}

@inproceedings{Tam25,
author = {Tamayo-Rios, Matthew and Schell, Alexander and Alaifari, Rima},
year = {2025},
pages = {},
title = {Scalable Signature Kernel Computations for Long Time Series via Local Neumann Series Expansions},
booktitle = {NeurIPS}
}

@article{Kra08,
  title={Emergent network topology at seizure onset in humans},
  author={Kramer, Mark and Kolaczyk, Eric and Kirsch, Hans },
  journal={Epilepsy Res.},
  volume={79},
  number={2--3},
  pages={173--186},
  year={2008},
}

@ARTICLE{Shu25,
  author={Shu, Qinji and Wang, Siyu and Ji, Feng and Feng, Hui and Hu, Bo},
  journal={IEEE Signal Process. Lett.}, 
  title={Finite-Size Convergence Bound of Graph Convolutional Networks by Graphon Analysis}, 
  year={2025},
  volume={32},
  number={},
  pages={3924-3928},
}

@inproceedings{Per14,
author = {Perozzi, Bryan and Al-Rfou, Rami and Skiena, Steven},
title = {{DeepWalk}: online learning of social representations},
year = {2014},
booktitle = {KDD},
pages = {701–710},
}

@ARTICLE{Ani16,
  author={Anis, Aamir and Gadde, Akshay and Ortega, Antonio},
  journal={IEEE Trans. Signal Process.}, 
  title={Efficient Sampling Set Selection for Bandlimited Graph Signals Using Graph Spectral Proxies}, 
  year={2016},
  volume={64},
  number={14},
  pages={3775-3789},
}

@ARTICLE{Cha18,
  author={Chamon, Luiz and Ribeiro, Alejandro},
  journal={IEEE Trans. Signal Process.}, 
  title={Greedy Sampling of Graph Signals}, 
  year={2018},
  volume={66},
  number={1},
  pages={34-47},
 }

@article{Sch23,
title = {A robustness analysis of blind source separation},
journal = {arXiv preprint arXiv:2303.10104},
year = {2023},
author = {Alexander Schell},
doi = {10.48550/arXiv.2303.10104},
url = {https://arxiv.org/abs/2303.10104},
}

@article{SchOb23,
  author  = {Schell, Alexander and Oberhauser, Harald},
  title   = {Nonlinear independent component analysis for discrete-time and continuous-time signals},
  journal = {Ann. Statist.},
  volume  = {51},
  number  = {2},
  pages   = {487--518},
  year    = {2023},
  doi     = {10.1214/23-AOS2256},
}

@article{She24,
title = {Sampling theory of jointly bandlimited time-vertex graph signals},
journal = {Signal Process.},
volume = {222},
pages = {109522},
year = {2024},
author = {Hang Sheng and Hui Feng and Junhao Yu and Feng Ji and Bo Hu},
}

@article{Ham11,
title = {Wavelets on graphs via spectral graph theory},
journal = {Appl. Comput. Harmon. Anal.},
volume = {30},
number = {2},
pages = {129-150},
year = {2011},
issn = {1063-5203},
author = {David Hammond and Pierre Vandergheynst and Rémi Gribonval},
}

@article{Zou25,
title = {Graph Neural Networks with scattering transform for network anomaly detection},
journal = {Eng. Appl. Artif. Intell.},
volume = {150},
pages = {110546},
year = {2025},
issn = {0952-1976},
author = {Abdeljalil Zoubir and Badr Missaoui},
}

@article{Zou20,
title = {Graph convolutional neural networks via scattering},
journal = {Appl. Comput. Harmon. Anal.},
volume = {49},
number = {3},
pages = {1046-1074},
year = {2020},
author = {Dongmian Zou and Gilad Lerman},
}

@book{van98,
  author    = {van der Vaart, A. W.},
  title     = {Asymptotic Statistics},
  year      = {1998},
  publisher = {Cambridge University Press},
}
